\documentclass[journal]{IEEEtran}

\usepackage{amsmath,graphicx}
\usepackage{algorithm,algorithmic}
\usepackage{cite,amssymb,amsfonts,color,textcomp}
\usepackage{bm}
\usepackage{balance}
\usepackage{booktabs}
\usepackage{amsthm}
\usepackage{setspace}
\usepackage{enumitem}

\def\thetabf{\boldsymbol \theta}

\def\rhobf{\boldsymbol \rho }

\def\gbf{{\bf g}}
\def\hbf{{\bf h}}

\def\nbf{{\bf n}}

\def\pbf{{\bf p}}

\def\sbf{{\bf s}}

\def\ubf{{\bf u}}
\def\vbf{{\bf v}}
\def\wbf{{\bf w}}
\def\xbf{{\bf x}}
\def\ybf{{\bf y}}
\def\zbf{{\bf z}}

\def\xbf{{\bf x}}
\def\ybf{{\bf y}}

\def\Cbf{{\bf C}}

\def\Ibf{{\bf I}}

\def\Bc{{\cal B}}

\def\Kc{{\cal K}}

\def\Pc{{\cal P}}

\def\Sc{{\cal S}}
\def\Tc{{\cal T}}

\def\Wc{{\cal W}}
\def\Xc{{\cal X}}
\def\Yc{{\cal Y}}

\def\ie{{\it i.e.,\ \/}}
\def\nn{\nonumber}

\def\Re{\mathfrak{R}\mathfrak{e}}
\def\Im{\mathfrak{I}\mathfrak{m}}
\def\etat{{\eta_{t}}}
\def\mae{{\mathbb{E}}}

\theoremstyle{definition}

\newtheorem{remark}{Remark}
\newtheorem{lemma}{Lemma}

\newtheorem{assumption}{Assumption}
\newtheorem{proposition}{Proposition}

\newtheorem{corollary}{Corollary}

\newenvironment{mylist}%
{\begin{list}{}%
    {%
      \setlength{\itemindent}{-5pt}%
      \setlength{\leftmargin}{12pt}%
      \setlength{\parsep}{\parskip}
      \setlength{\labelsep}{5pt}
      \setlength{\itemsep}{2pt}}}%
  {\end{list}}

\IEEEoverridecommandlockouts
\begin{document}

\title{\huge Improving Wireless Federated Learning via Joint Downlink-Uplink Beamforming over Analog Transmission}

\author{Chong~Zhang,~\IEEEmembership{Student~Member,~IEEE,}
        Min~Dong,~\IEEEmembership{Fellow,~IEEE,}
        Ben~Liang,~\IEEEmembership{Fellow,~IEEE,}\\
        Ali~Afana,~\IEEEmembership{Member,~IEEE,}
        and~Yahia~Ahmed%
        \thanks{Chong Zhang and Ben Liang are with the Department of ECE, University of Toronto, Canada (e-mail: \{chongzhang, liang\}@ece.utoronto.ca). Min Dong is with the Department of ECSE, Ontario Tech University, Canada (e-mail: min.dong@ontariotechu.ca). Ali Afana and Yahia Ahmed are with Ericsson Canada, Canada (e-mail: \{ali.afana, yahia.ahmed\}@ericsson.com).
        Part of this work was presented in \cite{Zhang&etal:WiOpt2023}.}
}%

\maketitle

\begin{abstract}
Federated learning (FL) over wireless networks using analog transmission can efficiently utilize the communication resource but is susceptible to errors caused by  noisy wireless links. In this paper, assuming a multi-antenna base station, we jointly design downlink-uplink beamforming  to maximize FL training convergence over time-varying wireless channels.
We derive the round-trip  model updating equation and use it to analyze the FL training convergence to capture the effects of downlink and uplink beamforming and the local model training on the global model update.
Aiming to maximize the FL training convergence rate, we
propose a low-complexity joint downlink-uplink beamforming (JDUBF) algorithm, which adopts a greedy approach to decompose the multi-round joint optimization and convert it into  per-round online joint optimization problems.
The per-round  problem is further decomposed into three subproblems over a
 block coordinate descent framework, where we show that each subproblem can be efficiently solved by projected gradient descent with fast  closed-form updates.
An  efficient initialization method
that leads to a closed-form initial point is also proposed to accelerate the convergence of JDUBF.
Simulation demonstrates that JDUBF  substantially  outperforms the conventional separate-link beamforming design.
\end{abstract}


\vspace*{-.5em}

\section{Introduction}
\label{sec:intro}

Federated learning (FL) is a widely recognized  method to train a machine learning model by multiple collaborating devices using their local training datasets\cite{Mcmahan&etal:2017}.
 A parameter server (PS) coordinates  the devices for local model updates and aggregates these updates to perform a global model update. In the wireless environment, the PS is usually hosted by a base station (BS), and FL requires
frequent exchange of a large amount of model parameters between the BS and many devices over the wireless links, stressing the limited communication resource, such as transmission bandwidth and power  \cite{Zhu&etal:2020}. Furthermore, the fluctuation of the wireless links and noisy reception at the receivers introduce distortion, resulting in training errors that degrade the FL performance in both training accuracy and convergence rate. Thus, it is crucial to obtain efficient communication design for wireless FL.

Many existing communication-efficient wireless FL solutions use digital transmission-then-aggregation schemes
for uplink acquisition of local parameters from the devices to BS
\cite{Du&etal:2020TSP, Yang&etal:2021,Xu&Wang:2021TWC,Amiri&etal:TWC2021,Wang&etal:JSAC2022}.
The  transmission and aggregation are designed separately in these schemes. Conventional digital transmission via orthogonal channels is used, which can consume large bandwidth and incur high latency when the number of devices is large.

To address this issue, analog transmission-and-aggregation schemes have been proposed and analyzed
\cite{Zhu&etal:TWC2020,Zhang&Tao:TWC2021,Wang&etal:ToN2024,Qu&etal:2023arxiv,Amiri&etal:TWC2022}.
In these schemes, devices  simultaneously transmit their local models via analog modulation over the shared multiple access channel,  achieving over-the-air  aggregation of local models by superposition.
Compared with the digital schemes, such analog schemes can significantly conserve the communication resource
and reduce communication latency.
The studies in \cite{Zhu&etal:TWC2020,Zhang&Tao:TWC2021,Wang&etal:ToN2024} focus on uplink aggregation while assuming an error-free downlink.
However, it is shown that the downlink transmission can be more vulnerable to  communication error than the uplink \cite{Qu&etal:2023arxiv}.
Noisy downlink transmission for wireless FL has been studied in \cite{Amiri&etal:TWC2022}
by assuming an error-free uplink, where it is shown that, since the gradient descent training method in FL is noise resilient,
analog transmission can be more efficient than digital transmission even for the downlink.

It is further recognized in \cite{Wei&Shen:2022,Guo&etal:JSAC2022,Shah&etal:2023JIOT,Wang&etal:JSAC2022b} that, especially with analog transmission, downlink and uplink communication for model parameter exchange  are coupled during the iterative FL\ training process. The noise and distortion  propagate over the FL communication and computation iterations. This suggests that a joint   downlink-uplink design is needed. However, the  intertwined process brings significant challenges
to tractable analysis and design optimization.
The literature on joint downlink-uplink communication design
for wireless FL is limited. A recent work has studied the effect of noisy downlink and uplink channels on the convergence of FL with non-independent and identically distributed local datasets  \cite{Wei&Shen:2022},
where a simple generic signal-in-noise receiver model is used to facilitate the analysis.
Analog designs have been proposed  for noisy  downlink and uplink
in both single-cell \cite{Guo&etal:JSAC2022,Shah&etal:2023JIOT} and multi-cell \cite{Wang&etal:JSAC2022b} scenarios.
However, these schemes  only consider single-antenna BSs, and
their solutions and convergence analysis cannot be applied to multi-antenna BSs, which are typical in practical wireless systems.

For  multi-antenna BSs, beamforming is an essential transmission technique that can be applied to enhance communication quality and reduce noise in wireless FL.
Receive beamforming for uplink analog over-the-air aggregation  is considered in   \cite{Chen&etal:IoT2018,Zhu&Huang:Globecom18}.
Subsequently, various uplink  beamforming designs have been proposed for analog schemes to improve the training performance of wireless FL
\cite{Yang&etal:TWC2020,Liu&etal:TWC2021,Kim&etal:TWC2023,Kalarde&etal:MSWiM2023,Kalarde&etal:2024,Zhang&etal:SPAWC24}.
These studies consider joint device transmit beamforming and BS receive beamforming.
It is shown that carefully  designed transmit and receive beamforming schemes can improve uplink over-the-air aggregation.
However, the existing literature mostly focus on uplink beamforming design for FL. Our recent work \cite{Zhang&etal:ICASSP2024} studies downlink beamforming in the context of multi-model FL. As far as we are aware, there is no existing work on joint downlink-uplink beamforming design, which is essential to optimize the overall learning performance.

Besides the analog over-the-air aggregation designs mentioned above, a few recent works have considered digital over-the-air aggregation
\cite{Zhu&etal:TWC2021,Sahin:TWC2024,Qiao&etal:JSAC2024}, where coding schemes with scalar or vector quantization schemes are proposed.
These studies focus on how to perform over-the-air computation using digital modulation and do not consider multi-antenna beamforming. The problems considered there  are beyond the scope of this paper.

\subsection{Contributions}

We consider  wireless FL between a multi-antenna BS and  collaborating devices, with noisy analog transmission over
both the downlink and the uplink and over-the-air aggregation  over the uplink.  We jointly design downlink-uplink beamforming  to optimize the communication process for  FL training over time-varying wireless channels.  Such a joint design is  challenging  as it requires the round-trip iterative model updating structure, and its dependency on the beamforming design can be highly complex.
We derive the round-trip global model updating equation, and use it to analyze the training convergence to capture the effects of both communication and the local model training on the global model update. Based on this,   we then propose a fast joint downlink-uplink
beamforming (JDUBF) algorithm to maximize the  FL learning performance.
The main contributions are summarized below:

\begin{mylist}
\item
We formulate the round-trip analog transmission and reception process with downlink and uplink beamforming. We utilize multicast beamforming \cite{Sidiropoulos&etal:TSP2006,Dong&Wang:TSP2020,Zhang&etal:TSP2023} for downlink broadcasting of the global model update to  devices, which is an  efficient beamforming technique to send common data to multiple devices simultaneously.
For the uplink, we address uplink beamforming for over-the-air aggregation in two cases based on the availability of channel state information (CSI) at the devices: 1) receive beamforming only, and 2) joint transmit-receive beamforming. Based on this downlink-uplink FL process, we  derive the overall  global model updating equation over each  communication round, capturing the effects of transceiver beamforming and  processing over noisy downlink-uplink  and the local model training on the global model update.

 \item
      Aiming at maximizing the FL training convergence rate, we use the obtained global updating equation  to formulate a problem of joint downlink-uplink beamforming and device power  optimization, to  minimize  the expected global training loss after $T$ communication rounds under transmit power constraints at the BS and the devices.
      For a tractable design, we derive upper bounds  on the global training loss   through FL training convergence analysis for the two uplink beamforming cases. The bounds   show the impact of  downlink-uplink beamforming and local device training on the convergence of
the global model update, through   a weighted sum of the inverse of received signal-to-noise ratios (SNRs) at the BS from all devices.

\item Based on the upper bounds, we propose a low-complexity JDUBF algorithm for each of the two uplink beamforming cases: JDUBF-R for  receive beamforming only, and JDUBF-TR for  joint transmit-receive beamforming.  The JDUBF algorithm adopts a greedy
approach to decompose the $T$-round joint beamforming and power allocation optimization problem into
separate per-round problems, each then solved in an online optimization manner based on the available global model update at the BS.
In particular,
we decompose each per-round joint optimization problem into three subproblems via the block coordinate descent (BCD) method \cite{Bertsekas:2016nonlinear}.
We show that each subproblem can be efficiently solved by the projected gradient descent (PGD) algorithm \cite{Levitin&Polyak:USSR1966} with fast closed-form gradient updates.
      To accelerate the convergence of the
      proposed methods, we also propose an efficient initialization method that uses closed-form initial points.

\item         Our simulation results under typical wireless network settings show that
      the proposed JDUBF algorithms substantially outperform the conventional  separate beamforming design over each link, leading to faster training convergence for a wide range of configurations of  the devices and the BS antennas.
      In particular, JDUBF-TR is shown to nearly attain the learning performance of ideal FL with
      error-free communication links. It outperforms
JDUBF-R at the cost of a higher communication overhead for required information for enabling transmit beamforming among devices.

\end{mylist}

\vspace*{-.5em}

\subsection{Organization and Notation}
The rest of this paper is organized as follows.
Section~\ref{sec:system_prob} presents the system model for wireless FL.
Section~\ref{sec:FL_alg} describes the downlink-uplink analog communication process for FL.
In Section~\ref{sec:prob_convrg}, we formulate the joint downlink-uplink beamforming optimization problem and develop the upper bounds  on the global training loss  via FL\ training convergence analysis.
Section~\ref{sec:joint_BF_alg} presents the low-complexity JDUBF-R and JDUBF-TR algorithms for the two uplink beamforming cases.
Simulation results are shown in Section~\ref{sec:simulations},
followed by the conclusion Section~\ref{sec:conclusion}.

\textit{Notation}:
Hermitian and transpose are denoted as
$(\cdot)^{\textsf{H}}$ and $(\cdot)^{\textsf{T}}$, respectively.
Real and imaginary parts
of a complex number are respectively denoted as $\Re{\{\cdot\}}$ and $\Im{\{\cdot\}}$.
The Euclidean norm of a vector
is denoted as $\|\cdot\|$,
the identity matrix is denoted as $\Ibf$,
and $\xbf \sim \mathcal{CN}({\bf{0}},{\bf{C}})$ denotes
a complex Gaussian random vector $\xbf$ with zero mean and covariance $\Cbf$.

\allowdisplaybreaks
\section{System Model}
\label{sec:system_prob}

\subsection{FL System}\label{sec:FL_system_intro}We consider a wireless FL system consisting of a server and $K$ devices
that collaboratively train a machine learning (ML) model.  Let  $\Kc = \{1, \ldots, K\}$ denote the set of devices. Each device $k \in \Kc$ holds a local training dataset with $S_k$ samples, denoted by
$\Sc_{k} = \{(\sbf_{k,i},v_{k,i}): 1 \le i \le S_k\}$, where
$\sbf_{k,i}\in\mathbb{R}^{b\times 1} $ is the $i$-th data feature vector, and $v_{k,i}$ is the label for this data sample.
Let $\thetabf\in\mathbb{R}^{D\times 1}$ denote the parameter vector of the ML model with $D$ parameters,
which predicts the true labels of data feature vectors.
Using their respective local training datasets, the devices collaboratively
train the global model $\thetabf$ at the server, while keeping their local datasets private.
The local training loss function representing the training error at device $k$ is defined as
\begin{align}
F_{k}(\thetabf)=\frac{1}{S_k}\sum_{i=1}^{S_k} L(\thetabf;\sbf_{k,i},v_{k,i}) \nn
\end{align}
where $L(\cdot)$ is the sample-wise training loss function associated with each data sample.
 The global training loss function is given by the weighted sum of the local loss functions over all $K$ devices:
\vspace*{-.5em}
\begin{align}
F(\thetabf) = \sum_{k=1}^{K}\frac{S_{k}}{S}F_{k}(\thetabf) \label{eq_global_local_equation}
\end{align}
where $S \triangleq  \sum_{k=1}^{K}S_{k}$ is the total number of training samples from all devices.
The learning objective is to find the optimal global model $\thetabf^\star$ that minimizes $F(\thetabf)$.

The devices communicate with the server over the downlink and uplink wireless channels to exchange the model update information iteratively  for model training. The training procedure in each communication round $t=0,1,\ldots$ is as follows:
\begin{itemize}
\item \emph{Downlink broadcast}: The server sends the parameter vector of the current global model $\thetabf_t$
to all $K$ devices via the downlink channels.
\item  \emph{Local model update}: Each device $k$ uses its dataset $\Sc_{k}$ to perform local training independently based on the received global model $\thetabf_t$. Specifically, device $k$ divides $\Sc_{k}$ into $J$ smaller mini-batches and then performs $J$ iterative local updates using the mini-batches to generate the updated local model $\thetabf^{J}_{k,t}$.
\item \emph{Uplink aggregation}: The $K$ devices send their updated local models $\{\thetabf^{J}_{k,t}\}_{k\in\Kc}$ to the  server  via the uplink  channels.
The server aggregates $\thetabf^{J}_{k,t}$'s to generate an updated global model $\thetabf_{t+1}$ for the next communication round $t+1$.
\end{itemize}

\subsection{Wireless Communication Model} Consider a wireless communication system where the server is
hosted by a BS equipped with $N$ antennas, and each device has a single antenna.
The BS applies downlink transmit beamforming to send the global model update $\thetabf_t$ to the $K$ devices and receive beamforming to the received signals from the $K$ devices to update the global model.

We consider analog communication to transmit the global and local models in the downlink and uplink, respectively. In particular, the BS and the devices send the respective values of  $\thetabf_t$ and $\{\thetabf^{J}_{k,t}\}_{k\in\Kc}$ directly under their transmit power budgets.
To use the communication bandwidth efficiently for uplink model aggregation,
we consider over-the-air computation via analog aggregation over the uplink multiple access channel. Specifically, the devices send their local models  $\thetabf^{J}_{k,t}$'s  to the BS simultaneously
over the same frequency band.
The BS receives the signal that is the superposition of $\thetabf^{J}_{k,t}$'s over the air. We assume that the control and signaling channels of the system still use digital transmissions and are perfect.

Due to the wireless downlink and uplink channels,  the received model updates at the devices and the BS are  the distorted noisy versions of  $\thetabf_t$ and $\{\thetabf^{J}_{k,t}\}_{k\in\Kc}$, respectively. These errors  in the model updates further propagate over the subsequent communication rounds, degrading the learning performance for FL model training.
 In this paper, we focus on this communication aspect of FL model training. Our goal is to  jointly design  downlink and uplink beamforming  to maximize the  learning performance of  FL over wireless transmission.

\section{Downlink-Uplink Analog Transmission for FL}
\label{sec:FL_alg}

To study the impact of the non-ideal wireless communication on the learning performance of FL, in this section,
we formulate the detailed transmission and reception process with  downlink and uplink beamforming
in the three stages of a communication round for FL model update as described in Section~\ref{sec:FL_system_intro}.

\subsection{Downlink Broadcast of Global Model Update}\label{subsec:dl_broadcast}
At the start of communication round $t$, the BS has the  current global model,  denoted by
$\thetabf_t = [\theta_{1,t}, \ldots, \theta_{D,t}]^{\textsf{T}}$.
For efficient transmission, we  represent $\thetabf_t$ using a complex signal vector,
with real and imaginary parts respectively containing the first and second halves of the elements in $\thetabf_t$.
Specifically,
we re-express $\thetabf_t = [(\tilde{\thetabf}_t^{\text{re}})^{\textsf{T}},
 (\tilde{\thetabf}_t^{\text{im}})^{\textsf{T}}]^{\textsf{T}}$, where
$\tilde{\thetabf}_t^{\text{re}} \triangleq [\theta_{1,t}, \ldots,
\theta_{\frac{D}{2},t}]^{\textsf{T}}$, and
$\tilde{\thetabf}_t^{\text{im}} \triangleq
[\theta_{\frac{D}{2}+1,t}, \ldots, \theta_{D,t}]^{\textsf{T}}$.
Let  $\tilde{\thetabf}_t$ denote the equivalent complex vector representation of $\thetabf_t$,
given by
$\tilde{\thetabf}_t = \tilde{\thetabf}^{\text{re}}_t + j\tilde{\thetabf}^{\text{im}}_t\in \mathbb{C}^{\frac{D}{2}\times 1}$.

Let $\hbf_{k,t}\in\mathbb{C}^{N\times 1}$ denote the downlink channel vector from the BS
to device $k\in\Kc$ in communication round $t$.
We assume $\{\hbf_{k,t}\}_{k\in\Kc}$ remain unchanged during the downlink transmission in round $t$ and are known perfectly at the BS.
For sending the common global model to all $K$ devices via beamforming, we consider multicast beamforming \cite{Sidiropoulos&etal:TSP2006,Dong&Wang:TSP2020,Zhang&etal:TSP2023}, which is an efficient transmission technique to send common signals to multiple devices simultaneously.
Specifically, let $\wbf^{\text{dl}}_t\in\mathbb{C}^{N\times 1}$ be the downlink multicast beamforming vector in communication round $t$.
The BS uses $\wbf^{\text{dl}}_t$ to send the common $\tilde{\thetabf}_t$ to all $K$ devices in $\frac{D}{2}$ channel uses.
Each device $k\in\Kc$ receives a complex signal vector, given by
\begin{align}
\ubf_{k,t} = (\wbf^{\text{dl}}_t)^{\textsf{H}}\hbf_{k,t}\tilde{\thetabf}_t + \nbf^{\text{dl}}_{k,t}  \nn
\end{align}
where
$\nbf^{\text{dl}}_{k,t}\in\mathbb{C}^{\frac{D}{2}\times 1}$ is  the receiver additive white Gaussian noise (AWGN)
vector with i.i.d. elements that are zero mean with variance $\sigma^2_{\text{d}}$.
The beamforming vector $\wbf^{\text{dl}}_t$ is subject to the BS transmit power budget. Let  $DP^{\text{dl}}$ be the total transmit power budget at the BS for sending the global model $\thetabf_t$ in $\frac{D}{2}$ channel uses, where $2P^{\text{dl}}$ represents the average transmit power budget per channel use for sending two elements of $\thetabf_t$ using a complex signal.
Then, for transmitting $\tilde{\thetabf}_t$, $\wbf^{\text{dl}}_t$ is subject to the  transmit power constraint $\|\wbf^{\text{dl}}_t\|^2\|\tilde{\thetabf}_t\|^2 \le DP^{\text{dl}}$.
The BS also sends the scaling factor
$\frac{\hbf^{\textsf{H}}_{k,t}\wbf^{\text{dl}}_t }
{|(\wbf^{\text{dl}}_t)^{\textsf{H}}\hbf_{k,t}|^2}$ to device $k$ via the downlink signaling channel to facilitate the receiver processing.
Device $k$  post-processes the received signal $\ubf_{k,t}$ using  this received scaling factor and obtains the estimate of the complex-valued global model update as
\begin{align}
\hat{\tilde{\thetabf}}_{k,t} & =
\frac{\hbf^{\textsf{H}}_{k,t}\wbf^{\text{dl}}_t }
{|(\wbf^{\text{dl}}_t)^{\textsf{H}}\hbf_{k,t}|^2}\ubf_{k,t} = \tilde{\thetabf}_t + \tilde{\nbf}^{\text{dl}}_{k,t}
\label{dl_device_signal}
\end{align}
where $\tilde{\nbf}^{\text{dl}}_{k,t} \triangleq
\frac{\hbf^{\textsf{H}}_{k,t}\wbf^{\text{dl}}_t }
{|(\wbf^{\text{dl}}_t)^{\textsf{H}}\hbf_{k,t}|^2}\nbf^{\text{dl}}_{k,t}$
is the post-processed noise vector at device $k$.
By the equivalence of real and complex signal representations between  $\thetabf_t$ and $\tilde{\thetabf}_t$,
device $k$ obtains the estimate of the global model $\thetabf_t$,
denoted by $\hat{\thetabf}_{k,t}$,
given by
\begin{align}
\hat{\thetabf}_{k,t} = \big[\Re{\big\{\hat{\tilde{\thetabf}}
_{k,t}\big\}^{\textsf{T}}}, \Im{\big\{\hat{\tilde{\thetabf}}_{k,t}
\big\}^{\textsf{T}}}\big]^{\textsf{T}}  = \thetabf_t + \hat{\nbf}^{\text{dl}}_{k,t}
\label{eq_dl}
\end{align}
where $\hat{\nbf}^{\text{dl}}_{k,t} \triangleq
[\Re{\{\tilde{\nbf}^{\text{dl}}_{k,t}\}^{\textsf{T}}},\,
\Im{\{\tilde{\nbf}^{\text{dl}}_{k,t}\}^{\textsf{T}}}]^{\textsf{T}}$ is the real-valued equivalent post-processed noise vector.

\subsection{Local Model Update}\label{subsec:device_update}
Device $k$ performs local model training based on $\hat{\thetabf}_{k,t}$ in \eqref{eq_dl} using its local dataset $\Sc_{k}$.
We assume each device adopts the widely used mini-batch stochastic gradient descent (SGD) algorithm to perform the local training
for minimizing the local training loss function $F_{k}(\thetabf)$  \cite{Bubeck:2015convex}.
It uses a subset of the training dataset to compute the gradient update at each iteration and achieves a favorable tradeoff between computational efficiency and convergence rate.
Assume $J$ mini-batch SGD iterations are used at each device for its local model update.
Let $\thetabf^{\tau}_{k,t} $ be the local model update by device $k$ at iteration $\tau \in \{0,\ldots,J-1\}$, with
$\thetabf^{0}_{k,t} = \hat{\thetabf}_{k,t}$, and let  $\Bc^{\tau}_{k,t}$ denote the mini-batch, \ie a subset of  $\Sc_{k}$,  at iteration  $\tau$.
Then, the local model update is given by
\begin{align}
\thetabf^{\tau+1}_{k,t} & = \thetabf^{\tau}_{k,t} - \eta_t \nabla F_{k}(\thetabf^{\tau}_{k,t}; \Bc^{\tau}_{k,t}) \nn\\
& = \thetabf^{\tau}_{k,t} - \frac{\eta_t}{|\Bc^{\tau}_{k,t}|}\sum_{(\sbf,v)\in\Bc^{\tau}_{k,t}}\!\!\!\nabla L(\thetabf^{\tau}_{k,t}; \sbf,v)
\label{eq_sgd}
\end{align}
where $\eta_{t}$ is the  learning rate in communication round $t$,
and $\nabla F_{k}$ and $\nabla L$ are the gradient functions  with respect to (w.r.t.) $\thetabf^{\tau}_{k,t}$.
After  $J$ iterations, device $k$ obtains the updated local model $\thetabf^{J}_{k,t}$.

\subsection{Uplink Aggregation of Local Model Updates}\label{subsec:ul_aggre}
The devices send their local model updates $\{\thetabf^{J}_{k,t}\}_{k\in\Kc}$ to
the BS using over-the-air aggregation to generate the global model update $\thetabf_{t+1}$ for the next communication round $t+1$.
For efficient transmission, similar to
the downlink, we represent $\thetabf^{J}_{k,t}$ using a complex vector with its real and imaginary parts containing half of the elements in $\thetabf^{J}_{k,t}$, respectively.
Let $\thetabf^{J}_{k,t} =
[(\tilde{\thetabf}^{J,\text{re}}_{k,t})^{\textsf{T}}, \,
(\tilde{\thetabf}^{J,\text{im}}_{k,t})^{\textsf{T}}]^{\textsf{T}}$,
where $\tilde{\thetabf}^{J,\text{re}}_{k,t}\triangleq[\theta^{J}_{k1,t}, \ldots, \theta^{J}_{k\frac{D}{2},t}]^{\textsf{T}}$
and $\tilde{\thetabf}^{J,\text{im}}_{k,t}
\triangleq[\theta^{J}_{k(\frac{D}{2}+1),t}, \ldots,
\theta^{J}_{kD,t}]^{\textsf{T}}$.
The equivalent complex vector representation of $\thetabf^{J}_{k,t}$ is thus given by $\tilde{\thetabf}^{J}_{k,t} = \tilde{\thetabf}^{J,\text{re}}_{k,t} + j\tilde{\thetabf}^{J,\text{im}}_{k,t}\in \mathbb{C}^{\frac{D}{2}\times 1}$.
Device $k$ transmits $\tilde{\thetabf}^{J}_{k,t}$ to the BS using a total of $\frac{D}{2}$ channel uses.

Let $\gbf_{k,t}\in\mathbb{C}^{N\times 1}$ denote the uplink channel vector from device $k\in\Kc$
to the BS in communication round $t$.
We assume $\{\gbf_{k,t}\}_{k\in\Kc}$ remain unchanged during the uplink transmission in round $t$ and are known perfectly at the BS.
Let $\tilde{\theta}^{J}_{kl,t}$ be the $l$-th element in $\tilde{\thetabf}^{J}_{k,t}$.
The devices send $\tilde{\theta}^{J}_{kl,t}$'s simultaneously
in channel use $l$.
As a result, the received signal vector at the BS for channel use $l$, denoted by $\vbf_{l,t}$, is given by
\begin{align}
\vbf_{l,t} = \sum_{k=1}^{K}\gbf_{k,t}a_{k,t}\tilde{\theta}^{J}_{kl,t} + \ubf^{\text{ul}}_{l,t} \nn
\end{align}
where $a_{k,t}\in\mathbb{C}$ is the  transmit beamforming weight at device $k$, and $\ubf^{\text{ul}}_{l,t}\in\mathbb{C}^{N\times 1}$ is the BS receiver AWGN vector with i.i.d. elements  that are zero mean
with variance $\sigma^2_{\text{u}}$.

The BS applies receive beamforming to  the received signal $\vbf_{l,t}$ over $N$ antennas, for $l=1,\ldots,\frac{D}{2}$.
Let $\wbf^{\text{ul}}_t\in\mathbb{C}^{N\times 1}$ be the unit-norm receive beamforming vector at the BS in communication round $t$, with $\|\wbf^{\text{ul}}_t\|^2 = 1$. Then, the post-processed received signal vector over all $\frac{D}{2}$ channel uses is given by
\begin{align}
\!\!\zbf_t  = \bigg(\!(\wbf^{\text{ul}}_t)^{\textsf{H}} \sum_{k=1}^{K}\gbf_{k,t}a_{k,t}\!\bigg)\tilde{\thetabf}^{J}_{k,t} + \nbf^{\text{ul}}_t   = \sum_{k=1}^{K}\alpha^{\text{ul}}_{k,t}\tilde{\thetabf}^{J}_{k,t} + \nbf^{\text{ul}}_t
\label{eq_beam_ul_overall}
\end{align}
where $\alpha^{\text{ul}}_{k,t} \triangleq
(\wbf^{\text{ul}}_t)^{\textsf{H}}\gbf_{k,t}a_{k,t}$ represents the effective uplink channel
from device $k$ to the BS after applying device transmit and BS receive beamforming,
and $\nbf^{\text{ul}}_t\in\mathbb{C}^{\frac{D}{2}\times 1}$ is the post-processed receiver noise with  the $l$-th element being
$(\wbf^{\text{ul}}_t)^{\textsf{H}}\ubf^{\text{ul}}_{l,t}$, for  $l=1,\ldots,\frac{D}{2}$.
The BS further scales $\zbf_t$  in \eqref{eq_beam_ul_overall} to obtain the complex equivalent global model update for the next round $t+1$:
\begin{align}
\tilde{\thetabf}_{t+1} = \frac{\zbf_t}{\sum_{k=1}^{K}\alpha^{\text{ul}}_{k,t}} = \sum_{k=1}^{K}\rho_{k,t}\tilde{\thetabf}^{J}_{k,t} + \tilde{\nbf}^{\text{ul}}_t \label{eq_global_update_0}
\end{align}
where $\rho_{k,t}\!  \triangleq\! \frac{\alpha^{\text{ul}}_{k,t}}{\sum_{j=1}^{K}\alpha^{\text{ul}}_{j,t}}$
represents the effective weight of device $k$ with $\sum_{k=1}^{K}\!\rho_{k,t}\! =\! 1$, and $\tilde{\nbf}^{\text{ul}}_t \!  \triangleq\! \frac{\nbf^{\text{ul}}_t}{\sum_{k=1}^{K}\alpha^{\text{ul}}_{k,t}}$  is the post-processed noise vector. The effective weight $\rho_{k,t}$ indicates the relative significance of device $k$'s local model update in the global model update. It reflects
the overall uplink processing effect including the device uplink transmission and BS receiver processing.

For transmitting the local model update $\thetabf^{J}_{k,t}$ at device $k$,
let $\tilde{\thetabf}^{J}_{k,t}$ be its equivalent complex representation and $\Delta\tilde{\thetabf}_{k,t} \triangleq \tilde{\thetabf}^{J}_{k,t} - \tilde{\thetabf}^{0}_{k,t}$ the corresponding model change after the local training.
Since $\thetabf^{0}_{k,t} = \hat{\thetabf}_{k,t}$, we have $\tilde{\thetabf}^{0}_{k,t} = \hat{\tilde{\thetabf}}_{k,t}$.
Then, from \eqref{dl_device_signal},
we express the global model $\tilde{\thetabf}_{t+1}$ in \eqref{eq_global_update_0} in terms of $\tilde{\thetabf}_{t}$
as
\begin{align}
\tilde{\thetabf}_{t+1} &   = \tilde{\thetabf}_t + \sum_{k=1}^{K}\rho_{k,t}\Delta\tilde{\thetabf}_{k,t}
+ \sum_{k=1}^{K}\rho_{k,t}\tilde{\nbf}^{\text{dl}}_{k,t} +  \tilde{\nbf}^{\text{ul}}_t.
\label{eq_global_update}
\end{align}
Finally, the real-valued  global model update $\thetabf_{t+1}$ is recovered from $\tilde{\thetabf}_{t+1}$ as
\begin{align}
\thetabf_{t+1}   =  [\Re{\{\tilde{\thetabf}_{t+1}\}^{\textsf{T}}},\,
\Im{\{\tilde{\thetabf}_{t+1}\}^{\textsf{T}}}]^{\textsf{T}}.  \nn
\end{align}

Note that the effective uplink channel $\alpha^{\text{ul}}_{k,t}$ in \eqref{eq_beam_ul_overall}
depends on the specific transmit/receive beamforming design of the uplink transmission.
In the following, we consider two cases: 1) receive beamforming only; 2) joint transmit-receive beamforming.

\subsubsection{ Receive Beamforming Only}\label{subsubsec:R_only}
We first consider the case where the BS does not send uplink CSI to the devices in order to keep the communication overhead low.
In this case, each device has no knowledge of CSI and only applies power scaling to its local model update for transmission, and the BS applies receive beamforming to post-process the received model parameters.
Specifically, device $k$ sets $a_{k,t} =\sqrt{p_{k,t}}$, where $p_{k,t}$ denotes the transmit power scaling factor for
device $k$ in round $t$.
After applying receive beamforming vector $\wbf^{\text{ul}}_t$ at the BS,
the resulting effective uplink channel $\alpha^{\text{ul}}_{k,t}$ in \eqref{eq_beam_ul_overall} is given by
\begin{align}
\alpha^{\text{ul}}_{k,t} = \sqrt{p_{k,t}}(\wbf^{\text{ul}}_t)^{\textsf{H}}\gbf_{k,t}, \ k \in \Kc. \label{eq_complex_imperfect_alpha}
\end{align}
Note that $\alpha^{\text{ul}}_{k,t}$ is complex-valued, as there is no transmit beamforming applied at the devices, and
thus, their signals received at the BS
cannot be perfectly phase aligned. This may significantly limit the quality of the aggregated local model updates in \eqref{eq_beam_ul_overall}.

\subsubsection{Joint Transmit-Receive Beamforming}
\label{subsubsec:joint_TR}
We also consider the case when CSI is available at the devices.
In this case, we can jointly design the device transmit beamforming weights $\{a_{k,t}\}_{k\in\Kc}$ and the BS receive beamforming vector $\wbf^{\text{ul}}_t$ to phase-align the effective uplink channels.
Specifically, the transmit beamforming weight $a_{k,t}$ at device $k$ is given by\footnote{The value of $a_{k,t}$ in \eqref{eq_device_weight} can be obtained at device $k$ by letting the BS send the scalar $\frac{\gbf_{k,t}^{\textsf{H}}\wbf^{\text{ul}}_t}
{|(\wbf^{\text{ul}}_t)^{\textsf{H}}\gbf_{k,t}|}$ to the device via the downlink signaling channel.
This requires additional communication overhead in each communication round.}
\begin{align}
a_{k,t} =\sqrt{p_{k,t}}  \frac{\gbf_{k,t}^{\textsf{H}}\wbf^{\text{ul}}_t}
{|(\wbf^{\text{ul}}_t)^{\textsf{H}}\gbf_{k,t}|}.
\label{eq_device_weight}
\end{align}
The effective uplink channel between device $k$ and the BS in this case is given by
\begin{align}
\alpha^{\text{ul}}_{k,t} = (\wbf^{\text{ul}}_t)^{\textsf{H}}\gbf_{k,t}a_{k,t}
 = \sqrt{p_{k,t}}|(\wbf^{\text{ul}}_t)^{\textsf{H}}\gbf_{k,t}|, \ k \in \Kc.
\label{eq_real_perfect_alpha}
\end{align}
All effective uplink channels are phase-aligned as the result of joint transmit-receive beamforming.

Finally, we note that each device is subject to transmit power budget.
Similar to the downlink, let $DP^{\text{ul}}_k$ be the total transmit power budget at device $k$ for sending each local model update $\thetabf^{J}_{k,t}$ in $\frac{D}{2}$ channel uses.
Then, for transmitting $\tilde{\thetabf}^{J}_{k,t}$, the transmit power constraint is $p_{k,t}\|\tilde{\thetabf}^{J}_{k,t}\|^2 \le DP^{\text{ul}}_k$.

\begin{remark}
The global model updating equation in \eqref{eq_global_update}  is derived from the entire round-trip FL procedure,
including downlink transmission via multicast beamforming, local model updates at devices via mini-batch SGD, and uplink over-the-air aggregation with a given transmit/receive beamforming scheme.
The second term in \eqref{eq_global_update} is a weighted average of local model changes at all devices.
It represents the aggregated local model change from $K$ devices as the result of uplink transmission.
The third term in \eqref{eq_global_update} is the downlink receiver noise from all $K$ devices aggregated at the BS via uplink transmission.
The fourth term in \eqref{eq_global_update} is the BS receiver noise from uplink transmission.
Overall, the global model updating equation \eqref{eq_global_update} is a noisy version of the aggregated local model updates. It shows how local model updates contribute to the global model update over the noisy communication channel and transmitter and receiver multi-antenna processing.
\end{remark}

\section{Joint Downlink-Uplink Beamforming for Wireless FL }
\label{sec:prob_convrg}

\subsection{Joint Downlink-Uplink Formulation}\label{subsec:prob_form}

Our objective in this paper is to design the communication aspect of FL,
to maximize the model training convergence rate.
In particular, we consider the expected global loss function after $T$ communication rounds to measure for training convergence rate.
Let $\pbf_t \triangleq [p_{1,t}, \ldots, p_{K,t}]^{\textsf{T}}$
contain the uplink transmit power scaling factors of all $K$
devices in round $t$.
Let $\wbf^{\text{dl}}\triangleq[(\wbf^{\text{dl}}_0)^{\textsf{H}},
\ldots, (\wbf^{\text{dl}}_{T-1})^{\textsf{H}}]^{\textsf{H}}\in\mathbb{C}^{TN\times 1}$,
$\wbf^{\text{ul}}\triangleq[(\wbf^{\text{ul}}_0)^{\textsf{H}},
\ldots, (\wbf^{\text{ul}}_{T-1})^{\textsf{H}}]^{\textsf{H}}\in\mathbb{C}^{TN\times 1}$,
and $\pbf\triangleq[\pbf^{\textsf{T}}_0, \ldots,
\pbf^{\textsf{T}}_{T-1}]^{\textsf{T}}\in\mathbb{R}^{TK\times 1}$
respectively denote the stacked downlink multicast beamforming vectors, the uplink receive beamforming vectors,
and the device power scaling vectors over the entire $T$ communication rounds.
We aim to find the joint beamforming and power control solution
$(\wbf^{\text{dl}},\wbf^{\text{ul}},\pbf)$ to maximize the expected global loss after $T$ communication rounds, which is formulated as
\begin{align}
\Pc_{o}: &\min_{\wbf^{\text{dl}},\wbf^{\text{ul}},\pbf} \ \mathbb{E}[F(\thetabf_T)]  \nn\\
\text{s.t.}& \quad \|\wbf^{\text{dl}}_t\|^{2}\|\thetabf_t\|^2 \leq D P^{\text{dl}}, \quad t\in \Tc, \label{constra_dl_T_timeslots} \\
&\quad  p_{k,t}\|\thetabf^{J}_{k,t}\|^2 \leq  D P^{\text{ul}}_k, \quad k\in \Kc,t\in \Tc, \label{constra_power_T_timeslots}\\
&\quad  \|\wbf^{\text{ul}}_t\|^2 = 1, \quad t\in \Tc \label{constra_ul_T_timeslots}
\end{align}
where $\Tc = \{0, \ldots, T-1\}$,
and $\mathbb{E}[\cdot]$ is the expectation taken w.r.t. the receiver noise at the devices and the BS and the mini-batch sampling for local training at each device. The constraints in  \eqref{constra_dl_T_timeslots} and \eqref{constra_power_T_timeslots} are the transmit power constraints at the BS and each device $k$, respectively.

Problem $\Pc_{o}$ is  a finite-horizon stochastic optimization problem. The expected global loss in the objective function is difficult to evaluate. Furthermore, the beamforming in both downlink and uplink is a one-to-many beamforming design, which is equivalent to a multicast beamforming problem that is nonconvex and NP-hard.
 To tackle this challenging problem, we first analyze the convergence rate of the global loss function
 and derive a more tractable upper bound on $\mathbb{E}[F(\thetabf_T)]$ as
 a function of the downlink and uplink beamforming design parameters $(\wbf^{\text{dl}},\wbf^{\text{ul}},\pbf)$.
We then develop our fast joint downlink-uplink beamforming algorithm to minimize this upper bound.

\vspace*{-.5em}
\subsection{Convergence Analysis on Global Training Loss}\label{subsec:convrg_bound}
The FL learning objective is to find the
optimal global model $\thetabf^{\star}$ that
minimizes the global training loss function $F(\thetabf)$.
Let $F^{\star}$ denote the minimum global loss
obtained by $\thetabf^{\star}$.
The expected optimality gap between the global loss under $\thetabf_T$ and the minimum global loss after $T$ communication rounds  is
$\mathbb{E}[F(\thetabf_T)] - F^{\star}$.
To  examine $\mathbb{E}[F(\thetabf_T)]$,
we can equivalently analyze $\mathbb{E}[F(\thetabf_T)]-F^{\star}$
based on the global model updating equation in \eqref{eq_global_update}.

We first make the following three assumptions on the local loss functions, the SGD, and the difference between the global and weighted average of the local loss gradients. These assumptions  are commonly
adopted for the convergence analysis of the FL model training \cite{Amiri&etal:TWC2022,Guo&etal:JSAC2022,Wang&etal:JSAC2022b}.

\begin{assumption}\label{assump_smooth}
The local loss functions $F_k(\cdot)$'s are differentiable and are  $L$-smooth:
$F_k(\ybf) \leq F_k(\xbf) + (\ybf - \xbf)^{\textsf{T}}\nabla F_k(\xbf) + \frac{L}{2}\|\ybf - \xbf\|^2$, $\forall~k\in\Kc$,
$\forall~\xbf, \ybf\in\mathbb{R}^{D\times 1}$.
Also, $F_k(\cdot)$'s are  $\lambda$-strongly convex:
$F_k(\ybf) \geq F_k(\xbf) + (\ybf - \xbf)^{\textsf{T}}\nabla F_k(\xbf) + \frac{\lambda}{2}\|\ybf - \xbf\|^2$, $\forall~k\in\Kc$,
$\forall~\xbf, \ybf\in\mathbb{R}^{D\times 1}$.
\end{assumption}

\begin{assumption}\label{assump_unbias}
The mini-batch SGD is unbiased: $\mathbb{E}_{\Bc}[\nabla F_{k}(\thetabf^{\tau}_{k,t}; \Bc^{\tau}_{k,t})] = \nabla F_{k}(\thetabf^{\tau}_{k,t})$, $\forall~k\in\Kc$, $\forall~t\in\Tc$, $\forall~\tau$.
The variance of the mini-batch stochastic local loss gradient is bounded by $\mu>0$:
For $\forall~k\in\Kc$, $\forall~t\in\Tc$, $\forall~\tau$,
$\mathbb{E}[\|\nabla F_{k}(\thetabf^{\tau}_{k,t}; \Bc^{\tau}_{k,t}) - \nabla F_{k}(\thetabf^{\tau}_{k,t})\|^2] \leq \mu$.
\end{assumption}

\begin{assumption}\label{assump_bound_diverg}
The gradient divergence is bounded by $\delta>0$:
For $\forall~k\in\Kc$, $\forall~t\in\Tc$,
$\mae[\| \nabla F(\thetabf_t) - \sum_{k=1}^{K}\phi_{k}\nabla F_{k}(\thetabf_t)  \|^2] \leq \delta$,
where $\phi_{k}\in\mathbb{R}$, and $\sum_{k=1}^{K}\phi_{k} = 1$.
\end{assumption}
In addition, we have the following assumption on the
gradient of the global loss and the change in the local model update.

\begin{assumption}\label{assump_bound_model}
The gradient of the global loss and the change in the local model update are finite:
For some $\zeta>0$, $\nu>0$, $\forall~k\in\Kc$, $\forall~t\in\Tc$,
$\|\nabla F(\thetabf_t)\|\leq \zeta$, $\|\Delta\thetabf_{k,t}\|\leq \nu$.
\end{assumption}

We point out that  $\|\nabla F(\thetabf_t)\|$ and $\|\Delta\thetabf_{k,t}\|$
being finite is typical for FL systems, and thus,
Assumption~\ref{assump_bound_model} generally holds.

From the global model updating equation in \eqref{eq_global_update},
we note that weight $\rho_{k,t}$ may be complex- or real-valued. It depends on the uplink beamforming case considered, as shown in
\eqref{eq_complex_imperfect_alpha} and \eqref{eq_real_perfect_alpha}
of Sections~\ref{subsubsec:R_only} and \ref{subsubsec:joint_TR}, respectively.
In the following convergence rate analysis,
we first assume $\rho_{k,t}$ takes the general form as a complex-valued scalar,
and then discuss the special case when $\rho_{k,t}$ is real-valued.

Based on Assumptions~\ref{assump_smooth}--\ref{assump_bound_model}, we now analyze the convergence rate of the expected global loss $\mathbb{E}[F(\thetabf_t)]$ for each round $t$
and provide an upper bound on $\mathbb{E}[F(\thetabf_T)] - F^{\star}$ after $T$ rounds.
Using the global model update obtained in \eqref{eq_global_update}, we first bound the expected change of the global loss function in two consecutive rounds as\vspace*{-.4em}
\begin{align}
\!\!& \mae[F(\thetabf_{t+1})-F(\thetabf_t)]  =
\sum_{k=1}^{K}\frac{S_{k}}{S}\mae[F_k(\thetabf_{t+1})-F_k(\thetabf_t)] \nn\\
\!\!& \stackrel{(a)}{\leq} \mae[(\thetabf_{t+1} - \thetabf_t)^{\textsf{T}} \nabla F(\thetabf_t)] + \frac{L}{2}\mae[\|\thetabf_{t+1} - \thetabf_t\|^2]  \label{eq_theta_diff_real}\\
\!\!& \stackrel{(b)}{=} \Re\{\mae[(\tilde{\thetabf}_{t+1}-
\tilde{\thetabf}_t)^{\textsf{H}} \nabla\tilde{F}(\thetabf_t)]\}+\frac{L}{2}\mae[\|\tilde{\thetabf}_{t+1}-\tilde{\thetabf}_t\|^2]\!
\nn \\
\!\!& \stackrel{(c)}{=}\!\Re\bigg\{\mae\bigg[
\bigg(\sum_{k=1}^{K}\rho_{k,t}\Delta\tilde{\thetabf}_{k,t}
\!+\!\sum_{k=1}^{K}\rho_{k,t}\tilde{\nbf}^{\text{dl}}_{k,t}\!+\!
\tilde{\nbf}^{\text{ul}}_t\bigg)^{\textsf{H}}\nabla \tilde{F}(\thetabf_t)\bigg]\bigg\}   \nn\\
&\quad+\frac{L}{2}\mae\bigg[\bigg\|\sum_{k=1}^{K}\rho_{k,t}\Delta\tilde{\thetabf}_{k,t}
\!+\!\sum_{k=1}^{K}\rho_{k,t}\tilde{\nbf}^{\text{dl}}_{k,t}\!+\!\tilde{\nbf}^{\text{ul}}_t\bigg\|^2\bigg] \label{eq_complex_theta_diff}
\end{align}
where
$(a)$ follows the $L$-smoothness of $F_k(\cdot)$'s in Assumption~\ref{assump_smooth} and the fact that $\nabla F(\thetabf) = \sum_{k=1}^{K}\frac{S_{k}}{S}\nabla F_{k}(\thetabf)$ from \eqref{eq_global_local_equation},
$(b)$ is based on an equivalent expression of \eqref{eq_theta_diff_real} by using the equivalent  complex representation    $\tilde{\thetabf}_t$ of $\thetabf_t$, where $\nabla \tilde{F}(\thetabf_t)$ denotes  the equivalent complex representation of the global loss gradient $\nabla F(\thetabf_t)$ in round $t$,
and $(c)$ is obtained following the global model update in \eqref{eq_global_update}.
The upper bound in \eqref{eq_complex_theta_diff} shows the effect of noisy channels and
beamforming processing at both downlink and uplink on the loss function.
Let $A_{1,t}$ denote the first term and
$B_{1,t}$ denote the expectation part $\mae[\cdot]$ in the
 second term in \eqref{eq_complex_theta_diff}, respectively. They are functions of the aggregated local model change, the joint downlink-uplink transmission processing, and the receiver noise at the BS and devices at round $t$.
Below, we provide upper bounds for $ A_{1,t}$ and $ B_{1,t}$, respectively.

We first derive the upper bound for $A_{1,t}$. For
$\Delta\tilde{\thetabf}_{k,t}$,
let $\Delta\thetabf_{k,t}\triangleq\thetabf^{J}_{k,t} - \thetabf^{0}_{k,t}$ denote the corresponding real-valued local model change after the local training at device $k$ in round $t$, where
$\Delta\thetabf_{k,t} =
[\Re{\{\Delta\tilde{\thetabf}_{k,t}\}^{\textsf{T}}},\,
\Im{\{\Delta\tilde{\thetabf}_{k,t}\}^{\textsf{T}}}]^{\textsf{T}}$.
Define
$\Delta\bar{\thetabf}_{k,t} =
[-\Im{\{\Delta\tilde{\thetabf}_{k,t}\}^{\textsf{T}}}, \,
\Re{\{\Delta\tilde{\thetabf}_{k,t}\}^{\textsf{T}}}]^{\textsf{T}}$.
Then, we have
\begin{align}
 A_{1,t}  & \stackrel{(a)}{=}\Re\bigg\{
 \mae\bigg[\bigg(\sum_{k=1}^{K}\rho_{k,t}\Delta\tilde{\thetabf}_{k,t}
\bigg)^{\textsf{H}}\nabla \tilde{F}(\thetabf_t)\bigg]\bigg\} \label{eq_A1_c}\\
&= \mae\bigg[\bigg(\sum_{k=1}^{K}\Re\{\rho_{k,t}\}\Delta\thetabf_{k,t}
\bigg)^{\textsf{T}}\nabla F(\thetabf_t)\bigg] \nn\\
& \qquad +\mae\bigg[\bigg(\sum_{k=1}^{K}\Im\{\rho_{k,t}\}\Delta\bar{\thetabf}_{k,t}
\bigg)^{\textsf{T}}\nabla F(\thetabf_t)\bigg]\label{eq_A1}
\end{align}
where $(a)$ is because the receiver noise vectors $\tilde{\nbf}^{\text{dl}}_{k,t}$'s at the devices and $\tilde{\nbf}^{\text{ul}}_t$ at the BS are zero mean and independent of $\nabla \tilde{F}(\thetabf_t)$,
and \eqref{eq_A1} is an equivalent expression of \eqref{eq_A1_c} by using the corresponding real-valued parameters.
Based on Assumptions~\ref{assump_smooth}--\ref{assump_bound_model},
we obtain an upper bound on $A_{1,t}$
in Lemma~\ref{lemma1}.

\begin{lemma}\label{lemma1}
 Consider the FL\ system described in Section~\ref{sec:FL_alg}.
 Let $\beta^{\text{re}}_{ t} \triangleq \sum_{k=1}^{K}\!|\Re\{\rho_{k,t}\}|$ and  $\beta^{\text{im}}_{ t} \triangleq \sum_{k=1}^{K}\!|\Im\{\rho_{k,t}\}|$.
 Let  $Q_t\triangleq1 - 4\eta_t^2J^2L^2$, and
 assume $\eta_tJ<\frac{1}{2L}$, $\forall~t\in\Tc$.
 Based on Assumptions~\ref{assump_smooth}--\ref{assump_bound_model}, $A_{1,t}$ is upper bounded as
\begin{align}
A_{1,t} & \leq  2\eta_tJ\bigg(  \frac{1-Q_t}{Q_t}  (\beta^{\text{re}}_{ t})^2 - \frac{1}{4}\bigg)\mae\left[\|\nabla F(\thetabf_t)\|^2\right] \nn\\
&\quad + \frac{D(1-Q_t)}{4\eta_tJQ_t}\beta^{\text{re}}_{ t}\sigma^2_{\text{d}}
\sum_{k=1}^{K}\frac{|\Re\{\rho_{k,t}\}|}{|(\wbf^{\text{dl}}_t)^{\textsf{H}}
\hbf_{k,t}|^2}  \nn\\
&\quad + \frac{\eta_tJ(1-Q_t)}{2Q_t}(4\delta + \mu)(\beta^{\text{re}}_{ t})^2  + \nu\zeta\beta^{\text{im}}_{t} + \eta_tJ\delta. \label{eq_A1_bound}
\end{align}
\end{lemma}
\IEEEproof
See Appendix~\ref{appA}.
\endIEEEproof

Note that by the definition of $\rho_{k,t}$ below \eqref{eq_global_update_0},
we have $\beta^{\text{re}}_{ t} \geq |\Re\{\sum_{k=1}^{K}
\rho_{k,t}\}|=1$, and also $\beta^{\text{im}}_{ t}\geq 0$.
For the special case of $\rho_{k,t}$ being
real-valued
(\ie in \eqref{eq_real_perfect_alpha} under the joint transmit-receive beamforming scheme),
we have $\beta^{\text{re}}_{t} =1$ and $\beta^{\text{im}}_{ t}= 0$.
Also, for the bound in \eqref{eq_A1_bound},
$\eta_t$ and $J$ are the parameters set in the SGD for
the local model update at each device,
$L, \mu, \delta$ are parameters specified in
Assumptions~\ref{assump_smooth}--\ref{assump_bound_diverg},
and $\nu, \zeta$ are parameters specified in
Assumption~\ref{assump_bound_model}.

For $B_{1,t}$, since the receiver noise at the BS is zero mean and independent of $\sum_{k=1}^{K}\rho_{k,t}(\Delta\tilde{\thetabf}_{k,t}+\tilde{\nbf}^{\text{dl}}_{k,t})$, we have
\begin{align}
\!\!\! B_{1,t} & =  \mae\bigg[\bigg\|\!\sum_{k=1}^{K}\!\rho_{k,t}(\Delta\tilde{\thetabf}_{k,t}\!+\!\tilde{\nbf}^{\text{dl}}_{k,t})\bigg\|^2\bigg]\!+\!\mae[\|\tilde{\nbf}^{\text{ul}}_t\|^2]\nn\\
\!\!\! & = \mae\bigg[\bigg\|\!\sum_{k=1}^{K}\!\rho_{k,t}(\Delta\tilde{\thetabf}_{k,t}\!+\!\tilde{\nbf}^{\text{dl}}_{k,t})\bigg\|^2\bigg] \!+\! \frac{D\sigma^2_{\text{u}}}{2\big|\!\sum^{K}_{k=1}\alpha^{\text{ul}}_{k,t}\big|^2}. \label{eq_B1t_firststep}
\end{align}
By Assumptions~\ref{assump_smooth}--\ref{assump_bound_model},
we upper bound $B_{1,t}$ in Lemma~\ref{lemma2}.
\begin{lemma}\label{lemma2}
Consider the FL\ system described in Section~\ref{sec:FL_alg}.
 Assume $\eta_tJ<\frac{1}{2L}$, $\forall~t\in\Tc$.
 Based on Assumptions~\ref{assump_smooth}--\ref{assump_bound_model},
$B_{1,t}$ is upper bounded as
\begin{align}
& B_{1,t} \!\leq\! \frac{2}{L^2}\bigg(\frac{1-Q_t}{Q_t}\bigg)
(\beta^{\text{re}}_{ t})^2\mae[\|\nabla F(\thetabf_t)\|^2]\! +\!\frac{D\sigma^2_{\text{u}}}{2\big|\!\sum^{K}_{k=1}\alpha^{\text{ul}}_{k,t}\big|^2} \nn\\
&\!+\!D\sigma^2_{\text{d}}\bigg(\!\frac{1-Q_t}{Q_t}
\beta^{\text{re}}_{t}\!
\sum_{k=1}^{K}\!\frac{|\Re\{\rho_{k,t}\}|}{|(\wbf^{\text{dl}}_t)^{\textsf{H}}
\hbf_{k,t}|^2}+\!\sum_{k=1}^{K}\!\frac{|\rho_{k,t}|^2}
{|(\wbf^{\text{dl}}_t)^{\textsf{H}}
\hbf_{k,t}|^2}\!\bigg) \nn\\
&\!\!+\!\frac{1-Q_t}{2L^{2}Q_t}\bigg(\Big(1-Q_t+\frac{Q_t}{J}\Big)\mu+ 4\delta\bigg)\!
(\beta^{\text{re}}_{ t})^2 \!+\!  2\nu^2(\beta^{\text{im}}_{ t})^2. \label{eq_B1t_lemma2}
\end{align}
\end{lemma}
\IEEEproof
See Appendix~\ref{appB}.
\endIEEEproof

We now  analyze the expected gap $\mathbb{E}[F(\thetabf_T)]-F^{\star}$
after $T$ communication rounds. From  \eqref{eq_complex_theta_diff},  the expected optimality gap  at round $t+1$ is bounded as
\begin{align}\label{theta_diff}
\mae[F(\thetabf_{t+1})]- F^{\star}\le \mae[F(\thetabf_t)]-
F^{\star}+ A_{1,t}+\frac{L}{2}B_{1,t}
\end{align}
where the RHS of \eqref{theta_diff} is further upper bounded
using \eqref{eq_A1_bound} and \eqref{eq_B1t_lemma2}. Summing up both sides over $t\in\Tc$ and rearranging the terms, we obtain the upper bound on $\mathbb{E}[F(\thetabf_T)] - F^{\star}$, which is stated in Proposition~\ref{thm:convergence}
below.

\begin{proposition}\label{thm:convergence}
Consider the FL\ system described in Section~\ref{sec:FL_alg}.
Let $V_t\triangleq\frac{1-Q_t+\sqrt{1-Q_t}}{Q_t}$,
and assume  $\frac{1}{2L(4(\beta^{\text{re}}_{ t})^2 \! + 1)}\le \eta_tJ<\frac{1}{2L}$, $\forall~t\in\Tc$. Based on Assumptions~\ref{assump_smooth}--\ref{assump_bound_model},
the expected gap $\mathbb{E}[F(\thetabf_T)] - F^{\star}$
after $T$ communication rounds is upper bounded by
\begin{align}
\mathbb{E}[ F(\thetabf_T)] \! - \! F^{\star} &  \!\leq \Gamma\!
\prod_{t=0}^{T-1}\!G_t + \Lambda + \!\!\sum_{t=0}^{T-2}\!
H(\wbf^{\text{dl}}_t, \wbf^{\text{ul}}_t, \pbf_t)\!\!\prod_{s=t+1}^{T-1}\!\!\!G_{s}\nn\\
& \qquad + H(\wbf^{\text{dl}}_{T-1}, \wbf^{\text{ul}}_{T-1}, \pbf_{T-1})
\label{eq_thm1}
\end{align}
where $\Gamma\triangleq\mathbb{E}[ F(\thetabf_0)]-F^{\star}$,
$\Lambda \triangleq\sum_{t=0}^{T-2}C_t \big(
\prod_{s=t+1}^{T-1}G_s\big)+ C_{T-1}$ with
\begin{align}
\!\!\!\!&G_t \triangleq \frac{1-Q_t}{4\eta_tJ \lambda}\big(4V_t(\beta^{\text{re}}_{ t})^2  -  1 \big) + 1,  \label{eq_prop1_complex_Gt}\\
\!\!\!\!&C_t \triangleq\frac{1-Q_t}{4L} \bigg(
V_t(4\delta+\mu) + 4\delta +\frac{\mu}{J} \bigg)
(\beta^{\text{re}}_{t})^2 + \eta_tJ\delta\nn\\
\!\!\!\!& \qquad\;\; +  L\nu^2\bigg(\beta^{\text{im}}_{t}+
\frac{\zeta}{L\nu}\bigg)\beta^{\text{im}}_{t}, \nn
\end{align}
and
\begin{align}
&H(\wbf^{\text{dl}}_t, \wbf^{\text{ul}}_t, \pbf_t) \triangleq
 \frac{LD}{2}\bigg(V_t\beta^{\text{re}}_{ t}\sigma^2_{\text{d}}\sum_{k=1}^{K}
 \frac{|\Re\{\rho_{k,t}\}|}{|(\wbf^{\text{dl}}_t)^{\textsf{H}}
 \hbf_{k,t}|^2}  \nn\\
 &\qquad\qquad +\sigma^2_{\text{d}}\sum_{k=1}^{K}\frac{|\rho_{k,t}|^2}
{|(\wbf^{\text{dl}}_t)^{\textsf{H}}
 \hbf_{k,t}|^2} + \frac{\sigma^2_{\text{u}}}{2\big|
 \sum^{K}_{k=1}\alpha^{\text{ul}}_{k,t}\big|^2}\bigg).
\label{eq_fc_Rt}
\end{align}
\end{proposition}
\begin{IEEEproof}
See Appendix~\ref{appC}.
\end{IEEEproof}

The upper bound on $\mathbb{E}[F(\thetabf_T)] - F^{\star}$ in  \eqref{eq_thm1}
shows how the  downlink-uplink transmission and the local device training impact the convergence of
the global model update.
In particular, the first term for $\Gamma$ shows the effect of  the  initial starting point $\thetabf_0$ on the convergence.
The second term $\Lambda$ is a weighted sum of
$C_t$'s over $T$ rounds. Each term for round $t$ in the sum accounts for
the  gradient difference of the local loss function from the optimal global loss $F^{\star}$ by using the mini-batch SGD at each device, and the uplink aggregation effect of the local gradient updates via beamforming.
The third and fourth terms represent a weighted sum of
$H(\wbf^{\text{dl}}_t, \wbf^{\text{ul}}_t, \pbf_t)$ over $T$
rounds. The expression of $H(\wbf^{\text{dl}}_t, \wbf^{\text{ul}}_t, \pbf_t)$ in \eqref{eq_fc_Rt} implicitly depends on $\wbf^{\text{ul}}_t$ and $\pbf_t$, since both $\rho_{k,t}$ and $\alpha^{\text{ul}}_{k,t}$ are functions of  $\wbf^{\text{ul}}_t$ and $\pbf_t$.

In particular, we note that $H(\wbf^{\text{dl}}_t, \wbf^{\text{ul}}_t, \pbf_t)$ in \eqref{eq_fc_Rt} is in the form of a weighted sum of the inverse of SNRs (\ie the noise-to-signal ratio). Two types of SNRs  are captured in $H(\wbf^{\text{dl}}_t, \wbf^{\text{ul}}_t, \pbf_t)$: 1) For the terms with  $\sigma_\text{d}^2$, the inverse of each term in the summation is related to the effective SNR
of the downlink multicast (due to downlink beamforming and receiver processing),
evaluated at the BS receiver (after uplink beamforming and receiver processing); 2) For the term with  $\sigma_\text{u}^2$, its inverse is the effective SNR of the uplink local model aggregation due to the device transmission scheme and receive beamforming and processing.

\vspace*{-.3em}

\subsubsection{Special case for joint transmit-receive beamforming}
\label{subsubsec:special_case_jointBF}
Proposition~\ref{thm:convergence} considers general weights
$\rho_{k,t}$'s for an arbitrary uplink beamforming case.
If we consider the joint transmit-receive beamforming scheme in
Section~\ref{subsubsec:joint_TR}, where the uplink transmission
can be phased aligned for local model aggregation as in
\eqref{eq_real_perfect_alpha}, $\rho_{k,t}$ becomes a positive
real-valued weight. In this case, we have $\beta^{\text{re}}_{ t} =1$ and $\beta^{\text{im}}_{ t}= 0$,
and $|\Re\{\rho_{k,t}\}| =  \rho_{k,t}$.
As a result, the bounds in Lemmas~\ref{lemma1} and \ref{lemma2} and Proposition~\ref{thm:convergence} are simplified as
follows.
\vspace*{-.4em}

\begin{lemma}\label{lemma3}
For $\rho_{k,t}$, $\forall~k\in\Kc$, $\forall~t\in\Tc$, being positive and real-valued,
and for $\eta_tJ<\frac{1}{2L}$, $\forall~t\in\Tc$,
the upper bound on $A_{1,t}$ in \eqref{eq_A1_bound} is given by
\begin{align}
& A_{1,t} \leq  2\eta_tJ\bigg(  \frac{1-Q_t}{Q_t}   - \frac{1}{4}\bigg)\mae\left[\|\nabla F(\thetabf_t)\|^2\right] \nn\\
&\qquad\qquad + \frac{D(1-Q_t)}{4\eta_tJQ_t}\sigma^2_{\text{d}}
\sum_{k=1}^{K}\frac{\rho_{k,t}}{|(\wbf^{\text{dl}}_t)^{\textsf{H}}
\hbf_{k,t}|^2}  \nn\\
&\qquad\qquad + \frac{\eta_tJ(1-Q_t)}{2Q_t}(4\delta + \mu)+ \eta_tJ\delta.
\nn
\end{align}
\end{lemma}
\vspace*{-.4em}
\begin{lemma}\label{lemma4}
For $\rho_{k,t}$, $\forall~k\in\Kc$, $\forall~t\in\Tc$, being positive and real-valued,
and for $\eta_tJ<\frac{1}{2L}$, $\forall~t\in\Tc$,
the upper bound on $B_{1,t}$ in \eqref{eq_B1t_lemma2} is given by
\begin{align}
& B_{1,t} \!\leq\! \frac{2}{L^2}\bigg(\frac{1-Q_t}{Q_t}\bigg)
\mae[\|\nabla F(\thetabf_t)\|^2]\! +\!\frac{D\sigma^2_{\text{u}}}{2\big|\!\sum^{K}_{k=1}\alpha^{\text{ul}}_{k,t}\big|^2} \nn\\
&\quad\;\;\;\!+D\sigma^2_{\text{d}}\bigg(\!\frac{1-Q_t}{Q_t}
\!
\sum_{k=1}^{K}\!\frac{\rho_{k,t}}{|(\wbf^{\text{dl}}_t)^{\textsf{H}}
\hbf_{k,t}|^2}+\!\sum_{k=1}^{K}\!\frac{\rho^2_{k,t}}
{|(\wbf^{\text{dl}}_t)^{\textsf{H}}
\hbf_{k,t}|^2}\!\bigg) \nn\\
&\quad\;\;\;\!+\frac{1-Q_t}{2L^{2}Q_t}\bigg(\Big(1-Q_t+\frac{Q_t}{J}\Big)\mu+ 4\delta\bigg). \nn
\end{align}
\end{lemma}
\vspace*{-.4em}
\begin{corollary}\label{thm:real_convergence}
For $\rho_{k,t}$, $\forall~k\in\Kc$, $\forall~t\in\Tc$, being positive and real-valued,
and for
$\frac{1}{10L}\le \eta_tJ<\frac{1}{2L}$, $\forall~t\in\Tc$, the upper bound on the expected  gap $\mathbb{E}[F(\thetabf_T)] - F^{\star}$
in \eqref{eq_thm1} is given by\begin{align}
\mathbb{E}[ F(\thetabf_T)] \! - \! F^{\star} &  \!\leq \Gamma\!
\prod_{t=0}^{T-1}\!G_t + \Lambda + \!\!\sum_{t=0}^{T-2}\!
H(\wbf^{\text{dl}}_t, \wbf^{\text{ul}}_t, \pbf_t)\!\!\prod_{s=t+1}^{T-1}\!\!\!G_{s}\nn\\
& \qquad + H(\wbf^{\text{dl}}_{T-1}, \wbf^{\text{ul}}_{T-1}, \pbf_{T-1})
\label{eq_prop2}
\end{align}
where
$G_t = \frac{1-Q_t}{4\eta_tJ \lambda}\big(4V_t  -  1 \big) + 1$,  
$C_t = \frac{1-Q_t}{4L} \big(
V_t(4\delta+\mu) + 4\delta +\frac{\mu}{J} \big)+ \eta_tJ\delta$,
and
\begin{align}
&H(\wbf^{\text{dl}}_t, \wbf^{\text{ul}}_t, \pbf_t) =
 \frac{LD}{2}\bigg(V_t\sigma^2_{\text{d}}\sum_{k=1}^{K}
 \frac{\rho_{k,t}}{|(\wbf^{\text{dl}}_t)^{\textsf{H}}
 \hbf_{k,t}|^2}  \nn\\
 &\qquad\qquad +\sigma^2_{\text{d}}\sum_{k=1}^{K}\frac{\rho_{k,t}^2}
{|(\wbf^{\text{dl}}_t)^{\textsf{H}}
 \hbf_{k,t}|^2} + \frac{\sigma^2_{\text{u}}}{2(\sum^{K}_{k=1}\!
 \alpha^{\text{ul}}_{k,t})^2}\bigg).
\label{eq_fc_Ht}
\end{align}
\end{corollary}

\begin{remark}
There are  different upper bounds on the expected gap $\mathbb{E}[ F(\thetabf_T)] -  F^{\star}$
that have been obtained in the FL literature \cite{Amiri&etal:TWC2022,Guo&etal:JSAC2022}. However, they are based on either idealized or simplified communication models without  considering multi-antenna processing.
Specifically,  assuming an error-free uplink, \cite{Amiri&etal:TWC2022} provides an upper bound for noisy downlink transmission using a single-antenna BS. The upper bound in \cite{Guo&etal:JSAC2022} is obtained by considering
joint noisy downlink-uplink transmission with a single-antenna BS. In contrast, the upper bounds we obtain for the FL convergence rate in \eqref{eq_thm1} and \eqref{eq_prop2} take into account the transmit and receive beamforming and processing over noisy downlink and uplink, which represent more realistic communication model for the practical multi-antenna systems than those in \cite{Amiri&etal:TWC2022,Guo&etal:JSAC2022}.
\end{remark}
\vspace*{-.3em}

Note that the upper bounds on $\mathbb{E}[F(\thetabf_T)] - F^{\star}$ given in Proposition~\ref{thm:convergence}
and Corollary~\ref{thm:real_convergence} are both in a more tractable form  than $\mathbb{E}[F(\thetabf_T)]$ in the original optimization problem $\Pc_{o}$, which can be explored
for the joint beamforming optimization design.
In the next section, we propose the joint downlink-uplink beamforming algorithms to minimize these two upper bounds directly.\vspace*{-.3em}
\vspace*{-.4em}
\section{Joint Downlink-Uplink Beamforming Design}
\label{sec:joint_BF_alg}
Instead of $\Pc_o$, we now optimize the joint downlink-uplink beamforming to minimize the expected optimality gap $\mathbb{E}[F(\thetabf_T)] - F^{\star}$. To do so, we consider the two uplink beamforming cases discussed in Sections~\ref{subsubsec:R_only} and \ref{subsubsec:joint_TR} for local model aggregation.

For the upper bound in \eqref{eq_thm1}, only
$H_{\text{}}(\wbf^{\text{dl}}_t,  \wbf^{\text{ul}}_t, \pbf_t)$ depends on the beamforming design. It is a weighted sum of the inverse of SNRs for downlink multicast and uplink aggregation,  as discussed below Proposition~\ref{thm:convergence}. In particular,
\begin{itemize}[leftmargin=*]
\item For the uplink receive beamforming only case:
$H(\wbf^{\text{dl}}_t,  \wbf^{\text{ul}}_t, \pbf_t)$ is given
in \eqref{eq_fc_Rt}. Substituting $\alpha^{\text{ul}}_{k,t}=\sqrt{p_{k,t}}
(\wbf^{\text{ul}}_t)^{\textsf{H}}\gbf_{k,t}$ from \eqref{eq_complex_imperfect_alpha}
into  $H(\wbf^{\text{dl}}_t,  \wbf^{\text{ul}}_t, \pbf_t)$, we arrive at the expression in \eqref{eq_fc_Rt_wBF}, denoted by $H^{\text{\tiny R}}(\wbf^{\text{dl}}_t, \wbf^{\text{ul}}_t, \pbf_t)$.

\item For the uplink joint transmit-receive beamforming case: $H(\wbf^{\text{dl}}_t,\!  \wbf^{\text{ul}}_t,\!\pbf_t)$ is given in
 \eqref{eq_fc_Ht}. Substituting $\alpha^{\text{ul}}_{k,t}= \sqrt{p_{k,t}}|(\wbf^{\text{ul}}_t)^{\textsf{H}}\gbf_{k,t}|$ from  \eqref{eq_real_perfect_alpha}
into $H(\wbf^{\text{dl}}_t,  \wbf^{\text{ul}}_t, \pbf_t)$, we arrive at the expression   in
\eqref{eq_fc_Ht_wBF}, denoted by $H^{\text{\tiny TR}}(\wbf^{\text{dl}}_t, \wbf^{\text{ul}}_t, \pbf_t)$.
\end{itemize}
\begin{figure*}[!htbp]
{\small
\begin{align}
& H^{\text{\tiny R}}(\wbf^{\text{dl}}_t, \wbf^{\text{ul}}_t, \pbf_t) \!=\!
\frac{LD}{2}\!
\displaystyle \Bigg(\!V_t \Bigg[\sum^{K}_{k=1}\Bigg|\Re\!\left\{\frac{\displaystyle
\sqrt{p_{k,t}}(\wbf^{\text{ul}}_t)^{\textsf{H}}\gbf_{k,t}}{\sum_{j=1}^{K}
\sqrt{p_{j,t}}(\wbf^{\text{ul}}_t)^{\textsf{H}}\gbf_{j,t}}\right\}\Bigg|\Bigg]
\sum^{K}_{k=1}
\frac{\sigma^{2}_{\text{d}}\bigg|\Re\bigg\{\frac{\sqrt{p_{k,t}}
(\wbf^{\text{ul}}_t)^{\textsf{H}}\gbf_{k,t} }
{\sum_{j=1}^{K}\sqrt{p_{j,t}}(\wbf^{\text{ul}}_t)^{\textsf{H}}
\gbf_{j,t}}\bigg\}\bigg|}{\big|(\wbf^{\text{dl}}_t)^{\textsf{H}}\hbf_{k,t}\big|^2}
\!+\! \frac{\displaystyle \sigma^{2}_{\text{d}}\sum^{K}_{k=1}\frac{p_{k,t}|(\wbf^{\text{ul}}_t)^{\textsf{H}}\gbf_{k,t}|^2}{|(\wbf^{\text{dl}}_t)^{\textsf{H}}\hbf_{k,t}|^2}\!+\!\frac{\sigma_\text{u}^2}{2}}
{\big|\sum_{k=1}^{K}\sqrt{p_{k,t}}(\wbf^{\text{ul}}_t)^{\textsf{H}}\gbf_{k,t}\big|^2}\Bigg)
\label{eq_fc_Rt_wBF} \\
&H^{\text{\tiny TR}}(\wbf^{\text{dl}}_t, \wbf^{\text{ul}}_t, \pbf_t) =
\frac{LD}{2}
\left(V_t\frac{\displaystyle \sigma^{2}_{\text{d}}\sum^{K}_{k=1}\frac{\sqrt{p_{k,t}}|(\wbf^{\text{ul}}_t)^{\textsf{H}}\gbf_{k,t}|}{|(\wbf^{\text{dl}}_t)^{\textsf{H}}\hbf_{k,t}|^2}}{\displaystyle\sum_{k=1}^{K}\sqrt{p_{k,t}}|(\wbf^{\text{ul}}_t)^{\textsf{H}}\gbf_{k,t}|} + \! \frac{\displaystyle \sigma^{2}_{\text{d}}\sum^{K}_{k=1}\frac{p_{k,t}|(\wbf^{\text{ul}}_t)^{\textsf{H}}\gbf_{k,t}|^2}{|(\wbf^{\text{dl}}_t)^{\textsf{H}}\hbf_{k,t}|^2}+\frac{\sigma_\text{u}^2}{2}}{\bigg(\displaystyle\sum_{k=1}^{K}\sqrt{p_{k,t}}|(\wbf^{\text{ul}}_t)^{\textsf{H}}\gbf_{k,t}|\bigg)^2}\right) \label{eq_fc_Ht_wBF} \\[.5em]
\hline \nn
\end{align}
}
\vspace*{-3em}
\end{figure*}
Define
\begin{align}
&\!\!\!\Psi(\wbf^{\text{dl}},\wbf^{\text{ul}},\pbf) \nn\\
&\!\!\!\!\triangleq\! \sum_{t=0}^{T-2}\!\!H(\wbf^{\text{dl}}_t,\!  \wbf^{\text{ul}}_t,\!\pbf_t)\!\!\!\prod_{s=t+1}^{T-1}\!\!
\! G_s\!  +\! H(\wbf^{\text{dl}}_{T-1},\! \wbf^{\text{ul}}_{T-1},\! \pbf_{T-1}), \label{eq_psi_expression}
\end{align}
which contains  $H(\wbf^{\text{dl}}_t,\!  \wbf^{\text{ul}}_t,\!\pbf_t)$'s in $T$ rounds. For both the upper bounds on
$\mathbb{E}[F(\thetabf_T)] - F^{\star}$ in \eqref{eq_thm1} and \eqref{eq_prop2}, omitting the first two constant terms,
we arrive at the joint optimization problem:
\begin{align*}
\Pc_{1}: & \min_{\wbf^{\text{dl}},\wbf^{\text{ul}},\pbf} \ \ \Psi_{\text{}}^{\text{}}(\wbf^{\text{dl}},\wbf^{\text{ul}},\pbf) \nn\\
& \qquad\text{s.t.} \quad \eqref{constra_dl_T_timeslots}\eqref{constra_power_T_timeslots}\eqref{constra_ul_T_timeslots}.
\end{align*}

\vspace*{-.8em}

\subsection{The JDUBF Algorithm}\label{sec:jdubf_alg}\vspace*{-.6em}
Problem $\Pc_{1}$ is a  $T$-horizon joint optimization problem involving $T$ rounds of the model updates.
In particular, $(\wbf^{\text{dl}}_t, \wbf^{\text{ul}}_t, \pbf_t)$ correlates with $(\wbf^{\text{dl}}_{t-1}, \wbf^{\text{ul}}_{t-1}, \pbf_{t-1})$  through the model updates $\|\thetabf_t\|^2$ and $\|\thetabf^{J}_{k,t}\|^2 $ in constraints \eqref{constra_dl_T_timeslots} and \eqref{constra_power_T_timeslots}, which is challenging to solve. Examining the objective function in  \eqref{eq_psi_expression},
we note that for both uplink beamforming cases above, if the condition on  $\eta_tJ$ in
Proposition~\ref{thm:convergence} or Corollary~\ref{thm:real_convergence} is satisfied, we have  $G_t > 0$, $\forall~t\in\Tc$, and thus, $\prod_{s=t+1}^{T-1}G_{s}> 0$.
Following this, we  separate
$\Psi(\wbf^{\text{dl}},\wbf^{\text{ul}},\pbf)$
into $T$ per-round objective functions and propose a greedy approach to minimize the per-round objective in each communication round $t$, given by
\begin{align}
\Pc_{2}^t: \min_{\wbf^{\text{dl}}_t, \wbf^{\text{ul}}_t, \pbf_t}&  H(\wbf^{\text{dl}}_t, \wbf^{\text{ul}}_t, \pbf_t) \nn\\
\text{s.t.}\ \ &  \|\wbf^{\text{dl}}_t\|^{2}\|\thetabf_t\|^2 \leq D P^{\text{dl}}, \label{constra_dl}\\
&   p_{k,t}\|\thetabf^{J}_{k,t}\|^2 \leq  D P^{\text{ul}}_k, \,\, k\in\Kc, \label{constra_power}\\
&  \|\wbf^{\text{ul}}_t\|^2 = 1 \label{constra_ul}
\end{align}
where  $H(\wbf^{\text{dl}}_t, \wbf^{\text{ul}}_t, \pbf_t)$ is given by either
\eqref{eq_fc_Rt_wBF} or \eqref{eq_fc_Ht_wBF}, depending on the uplink beamforming case considered.

Note that the solution $(\wbf^{\text{dl}}_t, \wbf^{\text{ul}}_t, \pbf_t)$ to problem $\Pc_{2}^t$ is computed at the BS at the beginning of communication round $t$ for the global model update in round $t$.
However, constraint \eqref{constra_power} contains the local model update  $\|\thetabf^{J}_{k,t}\|^2$, which is unavailable
at the beginning of round $t$.
To address this issue, we further propose to
solve $\Pc_{2}^t$ in an online optimization manner by replacing
 $\|\thetabf^{J}_{k,t}\|^2$ in \eqref{constra_power} with $\|\thetabf_t\|^2$, \ie the current global model
available at the BS at the beginning of round $t$, arriving at the following joint optimization problem
\begin{align}
\Pc_{3}^t: \min_{\wbf^{\text{dl}}_t, \wbf^{\text{ul}}_t, \pbf_t}&  H(\wbf^{\text{dl}}_t, \wbf^{\text{ul}}_t, \pbf_t) \nn\\
\text{s.t.}\ \ &  \eqref{constra_dl},\eqref{constra_ul},  \nn\\
&   p_{k,t}\|\thetabf_t\|^2 \leq  D P^{\text{ul}}_k, \,\, k\in\Kc. \label{const_power_Yt_delayed}
\end{align}

The expression of  $H(\wbf^{\text{dl}}_t, \wbf^{\text{ul}}_t, \pbf_t)$ given in either \eqref{eq_fc_Ht_wBF} or \eqref{eq_fc_Rt_wBF} is a complicated  nonconvex function of $(\wbf^{\text{dl}}_t, \wbf^{\text{ul}}_t, \pbf_t)$. Thus, finding the optimal solution to  $\Pc_3^t$ is challenging.  To compute a solution to $\Pc^t_3$ efficiently, we adopt the BCD method \cite{Bertsekas:2016nonlinear} to update  the downlink beamforming $\wbf^{\text{dl}}_t$ and uplink beamforming  $\wbf^{\text{ul}}_t$ and $\pbf_t$ alternatingly.
Let $\Wc^{\text{dl}}_t\triangleq \{\wbf^{\text{dl}}_t: \|\wbf^{\text{dl}}_t\|^{2}\|\thetabf_t\|^2 \leq D P^{\text{dl}}\}$,
$\Wc^{\text{ul}}_t\triangleq \{\wbf^{\text{ul}}_t: \|\wbf^{\text{ul}}_t\|^2 = 1\}$,
and $\Yc_t\triangleq \{\pbf_t: p_{k,t}\|\thetabf_t\|^2\leq  D P^{\text{ul}}_k, \,\, k\in\Kc\}$.
The BCD updating procedure to solve $\Pc_3^t$ is given as follows: At iteration $i$,
\begin{align}
\wbf^{\text{dl}(i+1)}_{t}= & \mathop{\arg\min}\limits_{\wbf^{\text{dl}}_t\in\Wc^{\text{dl}}_t}H(\wbf^{\text{dl}}_t,  \wbf^{\text{ul}(i)}_{t}, \pbf^{(i)}_{t}), \label{dl_BF_update_AO}\\
\wbf^{\text{ul}(i+1)}_{t}= & \mathop{\arg\min}\limits_{\wbf^{\text{ul}}_t\in\Wc^{\text{ul}}_t}H(\wbf^{\text{dl}(i+1)}_{t}, \wbf^{\text{ul}}_t,  \pbf^{(i)}_{t}), \label{ul_BF_update_AO}\\
\pbf^{(i+1)}_t= & \mathop{\arg\min}\limits_{\pbf_t\in\Yc_t}H_{\text{}}(\wbf^{\text{dl}(i+1)}_t, \wbf^{\text{ul}(i+1)}_t, \pbf_t). \label{power_update_AO}
\end{align}

Subproblems \eqref{dl_BF_update_AO}--\eqref{power_update_AO} are respectively a downlink beamforming problem, an uplink receive beamforming problem, and an uplink transmit power design problem.
Since the objective function of each subproblem is nonconvex with a complicated expression, we propose to apply PGD to solve each subproblem. PGD  \cite{Levitin&Polyak:USSR1966} is a first-order iterative algorithm that uses gradient updates to solve a  constrained minimization problem given by
$\min_{\xbf\in\Xc} f(\xbf)$, where $\Xc$ is the convex feasible set for $\xbf$.   Its updating procedure at iteration $j$ is given by
\begin{align}
\xbf_{j+1} = \Pi_{\Xc}\big(\xbf_j - \beta\nabla_{\xbf}f(\xbf_j)\big)
\label{PGD_update}
\end{align}
where $\beta > 0$ is the step size and $\Pi_\Xc(\xbf)$ denotes the projection of $\xbf$ onto $\Xc$.
PGD is guaranteed to find an approximate
stationary point for each subproblem in polynomial time \cite{Mokhtari&etal:NIPS2018}.

PGD can be applied to our problem, since the objective functions in \eqref{dl_BF_update_AO}--\eqref{power_update_AO} are differentiable and the feasible set of each subproblem is convex. Moreover, we adopt PGD because it is particularly suitable for subproblems \eqref{dl_BF_update_AO}--\eqref{power_update_AO}, where both the gradients and projections can be obtained in closed-form for fast update.
Specifically, the gradient of the objective function in each of subproblems \eqref{dl_BF_update_AO}--\eqref{power_update_AO} can be obtained in closed-form based on the expression in \eqref{eq_fc_Ht_wBF} or \eqref{eq_fc_Rt_wBF}. Also, the projection operation for each  subproblem is given by
\begin{itemize}[leftmargin=*]
\item  For subproblem \eqref{dl_BF_update_AO}:
\begin{align}
 \Pi_{\Wc^{\text{dl}}_t}(\wbf^{\text{dl}}_t)=
\begin{cases}
\wbf^{\text{dl}}_t   & {\text{if}\,\,\, \wbf^{\text{dl}}_t\in\Wc^{\text{dl}}_t}, \\
 \sqrt{\frac{D P^{\text{dl}}}{\|\wbf^{\text{dl}}_t\|^{2}\|\thetabf_t\|^2}}\wbf^{\text{dl}}_t & {\text{otherwise}}.
\end{cases}
\nn
\end{align}
\item For subproblem \eqref{ul_BF_update_AO}:
$\Pi_{\Wc^{\text{ul}}_t}(\wbf^{\text{ul}}_t)=\frac{\wbf^{\text{ul}}_t}{\|\wbf^{\text{ul}}_t\|}$.

\item For subproblem \eqref{power_update_AO}:
\[
[\Pi_{\Yc_t}(\pbf_t)]_k = \min\bigg\{p_{k,t}, \frac{D P^{\text{ul}}_k}
{\|\thetabf_t\|^2}\bigg\}, k\in\Kc
\]
where $[\Pi_{\Yc_t}(\pbf_t)]_k$ denotes the $k$-th element in  $\Pi_{\Yc_t}(\pbf_t)$.
\end{itemize}

In summary, our proposed JDUBF algorithm for $\Pc_{1}$
uses a greedy approach and solves the online version $\Pc_{3}^t$
per-round. It uses BCD \cite{Bertsekas:2016nonlinear} to solve
subproblems \eqref{dl_BF_update_AO}--\eqref{power_update_AO} alternatingly, and each is then solved via PGD \cite{Levitin&Polyak:USSR1966}.
The initial point of PGD for each subproblem is set to be the computed solution from the previous BCD iteration.  For example, for solving \eqref{dl_BF_update_AO} at BCD iteration $i$,
$\wbf^{\text{dl}(i)}_{t}$ is used as the initial point for
PGD to compute $\wbf^{\text{dl}(i+1)}_{t}$.

We denote JDUBF under the uplink receive beamforming only case as JDUBF-R,
and that under the uplink joint transmit-receive beamforming case as JDUBF-TR.
After obtaining the solution $(\wbf^{\text{dl}}_t, \wbf^{\text{ul}}_t, \pbf_t)$ to $\Pc^t_{3}$  in round $t$,  JDUBF-R and JDUBF-TR have different post-processing procedures at each device:
\begin{itemize}[leftmargin=*]
\item  JDUBF-R:  The benefit of receiver only beamforming is that the BS does not need to send any additional
information to devices, and thus  the
communication overhead is low.
In this case, each device directly sets  $a_{k,t}$ to its meet its power budget as  $a_{k,t} =  \frac{\sqrt{D P^{\text{ul}}_k}}
{\|\thetabf^{J}_{k,t}\|}$.

\item  JDUBF-TR: For device $k$ to apply the transmit beamforming weight $a_{k,t}$ in \eqref{eq_device_weight}, the BS computes $a_{k,t}$ based on  $(\wbf^{\text{ul}}_t, p_{k,t})$ and sends it via the downlink signaling channel to each device $k$. Device $k$ then recovers its power scaling solution $p_{k,t}=|a_{k,t}|$.
However,  since $p_{k,t}$ is computed at the BS using the global model update  $\thetabf_t$ in  \eqref{const_power_Yt_delayed},
applying it to the local model update $\thetabf^{J}_{k,t}$  may not satisfy the  transmit power constraint \eqref{constra_dl_T_timeslots}.
Each device $k$ needs to further adjust the power scaling factor
to ensure  the transmit power constraint  \eqref{constra_dl_T_timeslots} is met. Thus, the actual power
scaling factor used at device $k$ is determined by
$\tilde{p}_{k,t} = \min\big\{p_{k,t}, \frac{D P^{\text{ul}}_k}
{\|\thetabf^{J}_{k,t}\|^2}\big\}$.
Following this, the  transmit beamforming weight is updated as $a_{k,t}\leftarrow  \sqrt{\tilde{p}_{k,t}} \frac{a_{k,t}}{|a_{k,t}|}$, which yields
$a_{k,t} \! =\! \sqrt{\tilde{p}_{k,t}}\frac{\gbf_{k,t}^{\textsf{H}}\wbf^{\text{ul}}_t}
{|(\wbf^{\text{ul}}_t)^{\textsf{H}}\gbf_{k,t}|}$.
\end{itemize}

\emph{Computational Complexity:}
For JDUBF, the main computational complexity lies in computing the gradients of $H(\wbf^{\text{dl}}_t,  \wbf^{\text{ul}}_{t}, \pbf^{}_{t})$ in beamforming subproblems \eqref{dl_BF_update_AO}\eqref{ul_BF_update_AO} in each PGD iteration.
In particular, the required matrix-vector computation
for each beamforming gradient
requires $O(NK)$ flops. From our experiments,
for $N = 16\sim 64$ antennas and $K = 10\sim 20$ devices,
it typically takes $5\sim 1500$ iterations for PGD to converge.
We point out that since the updates are all in closed-form, the computation is fast despite it may take $1500$ iterations.
\vspace*{-.65em}
\subsection{The Initialization Method} \label{subsec:initial}\vspace*{-.35em}
JDUBF requires an  initial point
$(\wbf^{\text{dl}(0)}_{t},\wbf^{\text{ul}(0)}_{t},\pbf^{(0)}_{t})$
for the BCD iterations.  A good initial point  is desirable as it can facilitate  fast convergence. We propose an efficient initialization method that is based on separate downlink and uplink beamforming solutions.

\subsubsection{Downlink}
The downlink beamforming subproblem \eqref{dl_BF_update_AO} is a multicast beamforming problem for sending the global model  to all $K$ devices. The part of  the objective function $H(\wbf^{\text{dl}}_t, \wbf^{\text{ul}}_t, \pbf_t)$
in \eqref{eq_fc_Ht_wBF} or \eqref{eq_fc_Rt_wBF} w.r.t. $\wbf^{\text{dl}}_t$ is a summation term in the form of $\frac{\sigma^{2}_{\text{d}}}{|(\wbf^{\text{dl}}_t)^{\textsf{H}}\hbf_{k,t}|^2}$, \ie the inverse of received SNR at device $k$. Heuristically, the summation  would be dominated by the term with the minimum SNR. Thus, to minimize the summation, it is effective to maximize the minimum SNR, which is the MMF multicast beamforming problem \cite{Sidiropoulos&etal:TSP2006,Dong&Wang:TSP2020,Zhang&etal:TSP2023}.  Thus, we propose to use the beamforming solution to the MMF problem given by
\begin{align}
& \max_{\wbf^{\text{dl}}_t\in\Wc^{\text{dl}}_t}\min_{k\in\Kc} \,\,
|(\wbf^{\text{dl}}_t)^{\textsf{H}}\hbf_{k,t}|^2.  \label{dl_BF_update_SDUBF}
\end{align}
Note that the asymptotic multicast beamforming solution to the above problem, as  the number of antennas $N\to \infty $, is obtained in closed-form in (49) of \cite{Dong&Wang:TSP2020}. Thus, for fast initialization, we propose to use this asymptotic MMF multicast beamforming solution as  the initial  $\wbf^{\text{dl}(0)}_{t}$. It is in a form of MMSE beamformer with all parameters obtained in closed-form. We refer readers to   \cite{Dong&Wang:TSP2020} for the expression detail.


\subsubsection{Uplink}
For uplink subproblems \eqref{ul_BF_update_AO}\eqref{power_update_AO}, we notice in \eqref{eq_fc_Ht_wBF} or \eqref{eq_fc_Rt_wBF} that the term in   $H(\wbf^{\text{dl}}_t, \wbf^{\text{ul}}_t, \pbf_t)$  related to uplink aggregation
is an inverse to received SNR as
$\frac{\sigma_\text{u}^2/2}{(\sum_{k=1}^{K}\sqrt{p_{k,t}}|(\wbf^{\text{ul}}_t)^{\textsf{H}}\gbf_{k,t}|)^2}$.
We can further lower bound the SNR as  $\big(\sum_{k=1}^{K}\sqrt{p_{k,t}}|(\wbf^{\text{ul}}_t)^{\textsf{H}}\gbf_{k,t}|\big)^2\ge \sum_{k=1}^{K}p_{k,t}|(\wbf^{\text{ul}}_t)^{\textsf{H}}\gbf_{k,t}|^2$. Thus, we
 set $(\wbf^{\text{ul}(0)}_{t},\pbf^{(0)}_{t})$
as the solution of  the following uplink received SNR\ maximization problem:
\begin{align}\label{ul_init_BF_bound}
\max_{\wbf^{\text{ul}}_t\in\Wc^{\text{ul}}_t,\pbf_t\in \Yc_t}\frac{1}{\sigma_\text{u}^2}\sum_{k=1}^{K} p_{k,t}|(\wbf^{\text{ul}}_t)^{\textsf{H}}\gbf_{k,t}|^2.
\end{align}
The power solution to the above problem is the maximum power satisfying constraint \eqref{const_power_Yt_delayed}: $p_{k,t}=\frac{D P^{\text{ul}}_k}{\|\thetabf_t\|^2 }, \forall k$. The optimal solution $\wbf^{\text{ul}}_t$   to problem \eqref{ul_init_BF_bound}  is the eigenvector corresponding to the largest eigenvalue
of $\sum_{k=1}^{K}p_{k,t}\gbf_{k,t}\gbf^{\textsf{H}}_{k,t}$.

From the above proposed method, we obtain the initial point  $(\wbf^{\text{dl}(0)}_{t},\wbf^{\text{ul}(0)}_{t},\pbf^{(0)}_{t})$ all in closed-form.


\subsection{Separate Downlink and Uplink Beamforming Design}\label{sec:seperate_design}
To compare our joint beamforming approach JDUBF,  we also consider
the conventional method where downlink and uplink transmissions are designed separately for the communication system, which we refer to as  SDUBF. This method closely resembles the downlink and uplink beamforming problems we consider in our initialization method above:

\subsubsection{Downlink}
The downlink transmission of   the common global model to all $K$ devices is a multicast beamforming problem, which can be formulated as the MMF problem exactly as in \eqref{dl_BF_update_SDUBF} considered in the initialization method for our proposed method.  Instead of the asymptotic solution used there, we can solve the optimization problem more accurately by the projected subgradient algorithm (PSA) \cite{Zhang&etal:WCL2022}.

\subsubsection{Uplink}
For the uplink over-the-air aggregation, each device uses transmit beamforming weight $a_{k,t} =\sqrt{p_{k,t}}  \frac{\gbf_{k,t}^{\textsf{H}}\wbf^{\text{ul}}_t}
{|(\wbf^{\text{ul}}_t)^{\textsf{H}}\gbf_{k,t}|}$
in \eqref{eq_device_weight} to phase-align all $(\wbf^{\text{ul}}_t)^{\textsf{H}}\gbf_{k,t}$'s,
where $p_{k,t} = \frac{D P^{\text{ul}}_k}
{\|\thetabf^{J}_{k,t}\|^2}$
is the  maximum power scaling factor that meets
the power budget.
Assume all $p_{k,t}$'s are known perfectly at the BS. The uplink receive beamforming problem is
to maximize the received SNR of the aggregated signal, which is exactly the problem \eqref{ul_init_BF_bound}.
\vspace*{-.4em}

\section{Simulation Results}
\label{sec:simulations}

\begin{figure*}[t]
\centering
\vspace*{-1.7em}
\includegraphics[width=0.67\columnwidth]{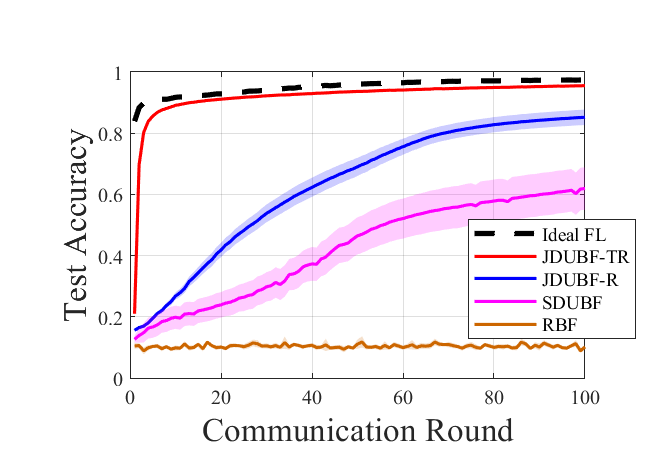}
\includegraphics[width=0.67\columnwidth]{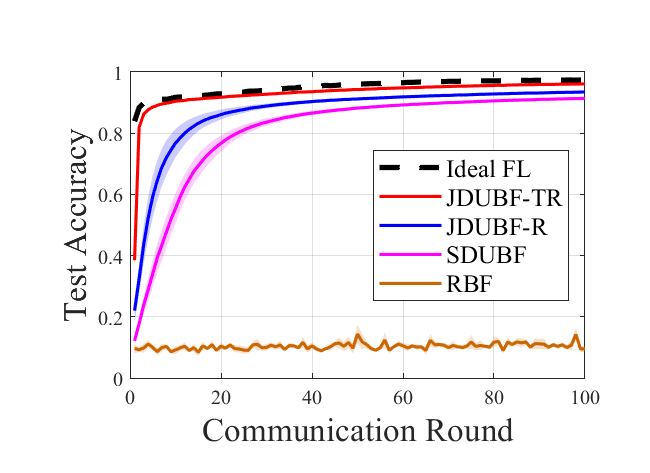}
\includegraphics[width=0.67\columnwidth]{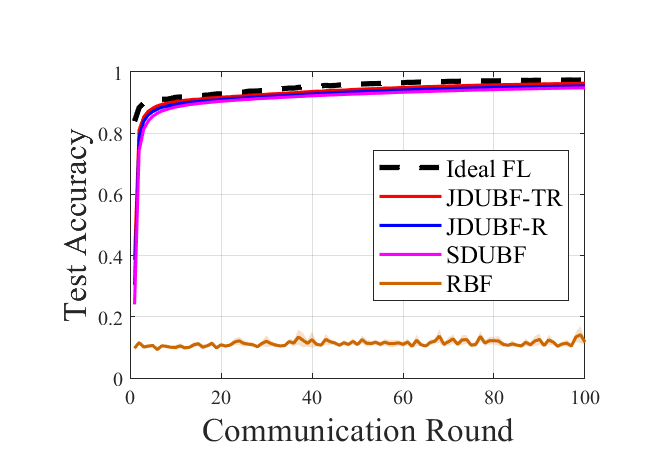}
\vspace*{-1.3em}\caption{Test accuracy vs. communication round $T$ with equal-distance devices for $(N,K) = (64,10)$.
Left: $G_{k} = -145~\text{dB}$, $\forall k$,  $(\text{SNR}_{\text{dl}},\text{SNR}_{\text{ul}}) = (-2,-10)$~dB.
Middle: $G_{k} = -140~\text{dB}$, $\forall k$,  $(\text{SNR}_{\text{dl}},\text{SNR}_{\text{ul}}) = (3,-5)$~dB.
Right: $G_{k} = -133~\text{dB}$, $\forall k$,  $(\text{SNR}_{\text{dl}},\text{SNR}_{\text{ul}}) = (10,2)$~dB.}
\label{Fig1:Acc_EqualDist_N64_K10} \vspace*{-.8em}
\end{figure*}

\subsection{Simulation Setup}
We consider the real-world dataset for image classification via a wireless FL system using typical 5G wireless specifications \cite{Yang&etal:5G2018}. We consider system  bandwidth  $10$~MHz and carrier frequency $2$~GHz.
The maximum transmit power at the BS is $47~\text{dBm}$, and that at each device is $23~\text{dBm}$. We assume the devices use $1$ MHz bandwidth for their uplink transmission.
The channels of all devices are generated i.i.d.\ as $\hbf_{k,t} =
\sqrt{G_{k}}\bar{\hbf}_{k,t}$ with
$\bar{\hbf}_{k,t}\sim\mathcal{CN}({\bf{0}},{\bf{I}})$ for downlink, and $\gbf_{k,t} = \sqrt{G_{k}}\bar{\gbf}_{k,t}$ with
$\bar{\gbf}_{k,t}\sim\mathcal{CN}({\bf{0}},{\bf{I}})$ for uplink, where  $G_k$ is the channel variance. We model $G_k$  based on the pathloss model:
$G_{k}[\text{dB}] = -161.3-35\log_{10}d_k - \psi_k$,
where $d_k$ is the BS-device distance in kilometers, and  $\psi_k$ is the shadowing random variable with standard deviation  $8~\text{dB}$.
Receiver noise power spectral density is $N_0 = -174~\text{dBm/Hz}$,
and noise figure is set to $N_F=8$~dB and $2$ dB at the device and  BS receivers, respectively.

We adopt the MNIST dataset \cite{Lecun&etal:MNIST} for  model training and testing.
MNIST consists of $6\times 10^4$ training samples and $1\times 10^4$ test samples from 10 different classes. Each sample
is a labeled image of size $28\times 28$ pixels with $\sbf_{k,i}\in\mathbb{R}^{784\times 1}$ and $v_{k,i}\in\{0,\ldots,9\}$ indicating the class.
We consider training a convolutional neural network with an $8\times3\times3$ ReLU convolutional layer,
a $2\times2$ max pooling layer, a ReLU fully-connected layer, and a softmax output layer,
resulting in $D=1.361\times 10^4$ model parameters in total.
We use the $ 10^4$ test samples to measure the test accuracy of the global model update $\thetabf_t$ at each round $t$. The training samples are randomly and evenly distributed  over $K$ devices, with the local dataset at device $k$ having $S_{k} = \frac{6\times 10^4}{K}$ samples.
For the local training via the SGD at each device, we set  $J=30$, the mini-batch size $|\Bc^{\tau}_{k,t}|=\frac{2\times 10^3}{K}, \forall k,\tau,t$, and the learning rate $\eta_t=0.1$, $\forall t$.
For both JDUBF-TR and JDUBF-R, we set the step size of PGD to $\beta=0.01$. 

\subsection{Performance Comparison}
We evaluate the performance of our  joint downlink-uplink beamforming design using the proposed JDUBF-TR and JDUBF-R method. We also consider the following three approaches for comparison:
\begin{itemize}[leftmargin=*]
\item \textbf{Ideal FL}\cite{Mcmahan&etal:2017}: FL\ with error-free downlink and uplink and perfect recovery of model parameters at the BS and devices. Specifically, it uses the global model update in \eqref{eq_global_update}, with receiver noise $\tilde{\nbf}^{\text{dl}}_{k,t} =  \tilde{\nbf}^{\text{ul}}_t = \bf{0}$
 and receiver post-processing weight $\rho_{k,t}=\frac{1}{K}$, $\forall k,t$.
This   benchmark provides the performance upper bound for all schemes.
\item \textbf{SDUBF}: The conventional separate downlink and uplink beamforming design based on SNR maximization, which is described in Section~\ref{sec:seperate_design}. The step size of PSA used in SDUBF is set to 0.01.
\item \textbf{RBF}: Randomly generated downlink and uplink beamforming vectors $\wbf^{\text{dl}}_t$ and $\wbf^{\text{ul}}_t$. The devices use the maximum transmit power and do not perform any transmit beamforming phase alignment.
This scheme is used to demonstrate the performance gain of  a properly designed beamforming solution.\end{itemize}

\begin{figure}[t]\vspace*{-1em}
\centering
\includegraphics[width=0.67\columnwidth]{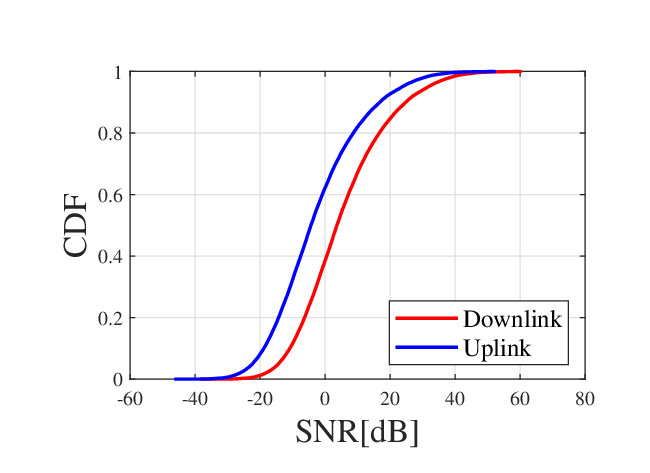}
\vspace*{-1.3em}\caption{The CDF of downlink and uplink SNRs with randomly located devices.}
\label{Fig:cdf_pathloss}
\vspace*{-.8em}
\end{figure}

\begin{figure*}[t]
\centering
\vspace*{-1.7em}
\includegraphics[width=0.67\columnwidth]{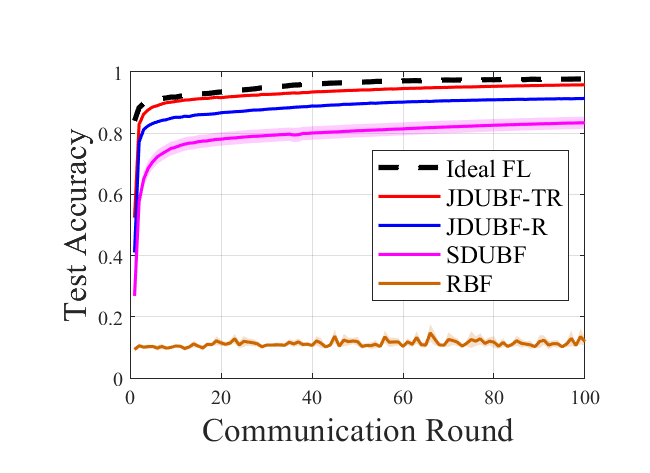}
\includegraphics[width=0.67\columnwidth]{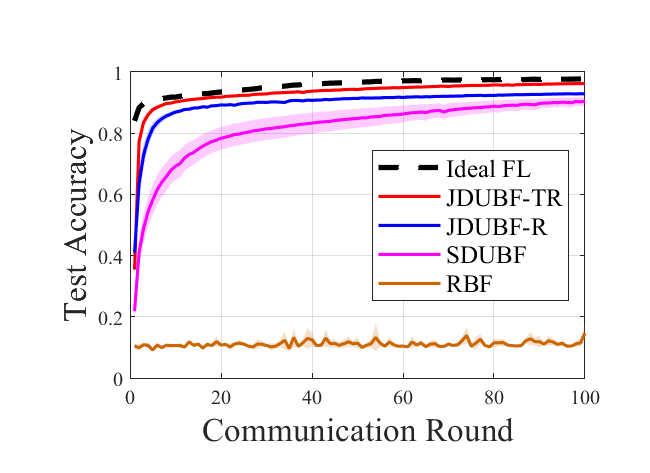}
\includegraphics[width=0.67\columnwidth]{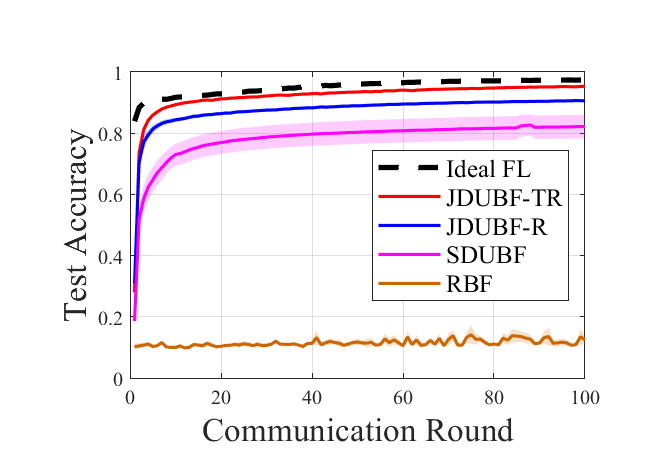}
\vspace*{-1.3em}\caption{Test accuracy vs. communication round $T$ with randomly located devices.
Left: $N = 16$,\,$K = 20$.
Middle: $N = 64$,\,$K = 20$.
Right: $N = 64$,\,$K = 10$.}
\label{Fig2:Acc_RandDist} \vspace*{-.8em}
\end{figure*}

\subsubsection{Performance with Equal-Distance Devices}
We first consider all devices have the same average channel quality  and study the test accuracy performance of different methods over communication round $T$. In particular, all devices are at the same distance $d_k$ from the BS. In this case, all device channels have the same pathloss $G_k$, and thus, the same downlink/uplink average nominal SNR  when the BS/device transmits with full power using a single antenna. The average nominal SNRs at  downlink and uplink are denoted by $\text{SNR}_{\text{dl}}$ and $\text{SNR}_{\text{ul}}$, respectively. We use them to indicate the level of channel quality in our experiments. We set $(N,K) = (64,10)$. All results are obtained by averaging over $20$ channel realizations.

Fig.~\ref{Fig1:Acc_EqualDist_N64_K10}-Left shows the test accuracy performance
for $(\text{SNR}_{\text{dl}},\text{SNR}_{\text{ul}}) = (-2,-10)$~dB,  a very low SNR environment on both links.
We also plot the 90\% confidence interval of each curve as the shadowed area.
We see that JDUBF-TR substantially outperforms the other schemes. In particular, it nearly attains the upper bound  under the Ideal FL for $T> 20$  and achieves an accuracy above $95\%$ for $T\approx 100$.
JDUBF-R is much worse than JDUBF-TR in this case. This is expected since JDUBF-R only uses uplink receive beamforming without  transmit beamforming at devices, and this results in much more noisy local model aggregation at a low SNR environment. Nonetheless, it still noticeably outperforms SDUBF, which employs uplink transmit-receive beamforming. This demonstrates the effectiveness of our joint downlink-uplink beamforming design for FL that considers both downlink and uplink communication effect for the global model update. In comparison, SDUBF has a much slower  training convergence rate, only reaching  $60\%$ test accuracy after  $T=100$.
For RBF, no training convergence is observed. This is because that RBF provides no beamforming gain, leading to highly suboptimal communication performance that affects the learning performance.

Fig.~\ref{Fig1:Acc_EqualDist_N64_K10}-Middle shows the test accuracy for $(\text{SNR}_{\text{dl}},\text{SNR}_{\text{ul}}) = (3~\text{dB},-5~\text{dB})$, a moderate downlink SNR\ environment and a low uplink SNR environment.
We see that as the channel quality improves, the learning performance improves, leading to a higher test accuracy under all beamforming design methods.
JDUBF-TR has the fastest convergence rate, nearly attaining the optimal performance after $10$ rounds, while JDUBF-R and SDUBF approach the upper bound  with slower convergence rates
and are worse than JDUBF-TR after 100 rounds.
JDUBF-R still converges faster than SDUBF with a slightly better accuracy after 100 rounds.
Fig.~\ref{Fig1:Acc_EqualDist_N64_K10}-Right shows the case for $(\text{SNR}_{\text{dl}},\text{SNR}_{\text{ul}}) = (10~\text{dB},2~\text{dB})$, a high downlink SNR environment and a moderate uplink SNR environment.
We see that the performance of all  three methods JDUBF-TR, JDUBF-R and SDUBF  are close  to the Ideal FL, in terms of both convergence rate and the achieved test accuracy.  This is expected as when the device channel quality improves, the impact of noisy communication on the performance reduces, and the different beamforming design methods can all lead to good learning performance close to the Ideal FL.

\subsubsection{Performance with Randomly Located Devices}

We now consider randomly located devices within the cell radius, where the BS-device distance    $d_k \in(0.02\text{~km}, 0.5\text{~km})$ for each device. We study the test accuracy performance of the proposed methods over communication round $T$ under different settings of $(N,K)$.
All results are obtained by averaging over 5 drops of device locations, each with $20$ channel realizations. We first show the cumulative distribution functions (CDFs) of $\text{SNR}_{\text{dl}}$ and $\text{SNR}_{\text{ul}}$ among devices at donwlink and uplink, respectively, in Fig.~\ref{Fig:cdf_pathloss}.
It shows that $\text{SNR}_{\text{dl}}$ is in the range of $(-20~\text{dB},40~\text{dB})$.     $\text{SNR}_{\text{ul}}$ at uplink is in the range of $(-30~\text{dB},35~\text{dB})$, about $5\sim 7$ dB worse than downlink. This is  expected due to different maximum transmit power at the BS and devices.

Fig.~\ref{Fig2:Acc_RandDist}-Left shows the test accuracy performance
 for $(N,K) = (16,20)$. We see that JDUBF-TR converges fast and nearly attains the upper bound  under the Ideal FL after $20$ communication rounds. It achieves an accuracy above $95\%$ at $100$ rounds. JDUBF-R is  worse than JDUBF-TR without uplink transmit beamforming, but it still achieves above $90\%$ accuracy at 100 rounds.
SDUBF has a much slower  training convergence rate. After 100 rounds,
it reaches test accuracy between  $80\%-85\%$.
RBF again performs poorly without beamforming gain. No training convergence is observed for RBF,
which leads to an accuracy  $\sim 10\%$ for all rounds.

Fig.~\ref{Fig2:Acc_RandDist}-Middle shows the case for $(N,K) = (64,20)$. Compared with those in Fig.~\ref{Fig2:Acc_RandDist}-Left,
both JDUBF-R and SDUBF  have noticeable performance improvement due to   more BS antennas leading to higher beamforming again at  downlink and uplink for improved wireless FL communication.  This results in improved overall learning performance.
The further performance improvement of JDUBF-TR is less noticeable, as it is already close to the Ideal FL for $N=16$ in
Fig.~\ref{Fig2:Acc_RandDist}-Left.

Fig.~\ref{Fig2:Acc_RandDist}-Right shows the test accuracy for $(N,K) = (64,10)$.
 Comparing Figs.~\ref{Fig2:Acc_RandDist}-Middle and Right, we see that as $K$ reduces from $20$ to $10$, the change of test accuracy of JDUBF-TR and JDUBF-R is hardly noticeable. However, that of SDUBF has a more significant reduction. The performance depredation is due to fewer devices resulting in less (distributed) transmit beamforming gain for uplink over-the-air aggregation.
 The learning performance of SDUBF under separate downlink and uplink design is more sensitive to the number of participating devices for aggregation.  In summary,
from Fig.~\ref{Fig2:Acc_RandDist} with different $N$ and $K$ values, we see that JDUBF-TR and JDUBF-R based on joint downlink-uplink beamforming design  are effective methods to combat noisy and imperfect communication links to facilitate wireless FL under various system configurations. They achieve fast training convergence and high test accuracy.

\section{Conclusion}
\label{sec:conclusion}
 This paper considers the transceiver beamforming design  for wireless FL experiencing fluctuated wireless links and noisy reception that degrade the learning performance. We propose a joint downlink-uplink beamforming design approach to maximize the FL training convergence over such  a wireless environment.
We first obtain
 the  round-trip global model updating equation, which captures the impact of transmitter and receiver processing, noisy reception, and local model training. These factors are then reflected in the upper bound we derive on the expected global training loss after $T$ rounds, and we formulate the joint downlink-uplink beamforming optimization problem  to minimize this upper bound after $T$ rounds.
Depending on whether the knowledge of CSI is available at the devices, we propose two  joint downlink-uplink beamforming methods,
JDUBF-TR and JDUBF-R. Each method uses a greedy approach to solve a per-round joint online optimization problem, which is further decomposed into three subproblems that are solved alternatingly. PGD is used for each subproblem, which yields fast closed-from updates. We also propose a closed-form initialization method to accelerate the fast convergence of JDUBF-TR and JDUBF-R. Simulation results  show that
our proposed methods substantially outperform the conventional separate-link-based beamforming design for various number of antennas and devices. In particular, JDUBF-TR nearly attains the learning performance of ideal FL with error-free communication links.

\begin{appendices}
\section{Proof of Lemma~\ref{lemma1}}\label{appA}\vspace*{-.2em}
\IEEEproof
For bounding $A_{1,t}$, we separately bound the two terms in \eqref{eq_A1}.
Specifically, let $A_{2,t}$ and $A_{3,t}$ denote the first and second terms in \eqref{eq_A1}, respectively.
For $A_{2,t}$, we have
\vspace*{-.5em}
\begin{align}
\!& A_{2,t} \stackrel{(a)}{=} -\eta_t\sum_{\tau=0}^{J-1}\mae\bigg[\bigg(\sum_{k=1}^{K}\Re\{\rho_{k,t}\}\nabla F_{k}(\thetabf^{\tau}_{k,t})\bigg)^{\textsf{T}}\!\!\nabla F(\thetabf_t)\bigg]\nn\\
\!& \stackrel{(b)}{\leq} -\frac{\eta_t}{2}\sum_{\tau=0}^{J-1}\mae\left[\|\nabla F(\thetabf_t)\|^2\right]\nn\\
\!& +\frac{\eta_t}{2}\sum_{\tau=0}^{J-1}\mae\bigg[\bigg\|\nabla F(\thetabf_t) - \sum_{k=1}^{K}\Re\{\rho_{k,t}\}\nabla F_{k}(\thetabf^{\tau}_{k,t})\bigg\|^2\bigg]\nn\\
\!&  \stackrel{(c)}{\leq} -\frac{\eta_tJ}{2}\mae\left[\|\nabla F(\thetabf_t)\|^2\right] \nn\\
\!& + \eta_t\sum_{\tau=0}^{J-1}\mae\bigg[\bigg\|\nabla F(\thetabf_t) - \sum_{k=1}^{K}\Re\{\rho_{k,t}\}\nabla F_{k}(\thetabf_t)\bigg\|^2\bigg]\nn\\
\!& + \eta_t\sum_{\tau=0}^{J-1}\mae\bigg[\bigg\|\sum_{k=1}^{K}\Re\{\rho_{k,t}\}(\nabla F_k(\thetabf_t) - \nabla F_{k}(\thetabf^{\tau}_{k,t}))\bigg\|^2\bigg]\nn\\
\!&  \stackrel{(d)}{\leq}  -\frac{\eta_tJ}{2}\mae\left[\|\nabla F(\thetabf_t)\|^2\right] + \eta_tJ\delta \nn\\
\!& +\! \eta_t(\beta^{\text{re}}_{ t})^2\!\sum_{\tau=0}^{J-1}\!\mae\bigg[\bigg\|\!\sum_{k=1}^{K}\!\frac{|\Re\{\rho_{k,t}\}|}{\beta^{\text{re}}_{ t}}\!\left\|\nabla F_k(\thetabf_t)\!-\!\!\nabla F_{k}(\thetabf^{\tau}_{k,t})\right\|\!\bigg\|^2\bigg]\nn\\
\!& \stackrel{(e)}{\leq} -\frac{\eta_tJ}{2}\mae\left[\|\nabla F(\thetabf_t)\|^2\right] + \eta_tJ\delta \nn\\
\!& +\! \eta_t\beta^{\text{re}}_{ t}\!\sum_{k=1}^{K}\!|\Re\{\rho_{k,t}\}|
\sum_{\tau=0}^{J-1}\!\mae\left[\left\|\nabla F_k(\thetabf_t)\!-\!\nabla F_{k}(\thetabf^{\tau}_{k,t})\right\|^2\right]  \nn\\
\!& \stackrel{(f)}{\leq} -\frac{\eta_tJ}{2}\mae\left[\|\nabla F(\thetabf_t)\|^2\right] + \eta_tJ\delta \nn\\
\!& + \eta_tL^2\beta^{\text{re}}_{ t}\!\sum_{k=1}^{K}\!|\Re\{\rho_{k,t}\}|\sum_{\tau=0}^{J-1}\mae\left[\|\thetabf_t- \thetabf^{\tau}_{k,t}\|^2\right]
\label{eq_A2t_sum}
\end{align}
where $(a)$ follows $\Delta\thetabf_{k,t} = -\eta_t\sum_{\tau=0}^{J-1} \nabla F_{k}(\thetabf^{\tau}_{k,t}; \Bc^{\tau}_{k,t})$
and the unbiasedness of the mini-batch SGD from Assumption~\ref{assump_unbias},
$(b)$ and
$(c)$ are based on $\|\xbf + \ybf\|^2\leq 2\|\xbf \|^2 + 2\|\ybf\|^2, \forall~\xbf,\ybf$,
$(d)$ follows Assumption~\ref{assump_bound_diverg},
$(e)$ is based on the Jensen's inequality,
and $(f)$ follows the $L$-smoothness of  $F_k(\cdot)$ from Assumption~\ref{assump_smooth}.
Let $A_{4,t}$ denote  $\sum_{\tau=0}^{J-1}\mae[\|\thetabf_t-\thetabf^{\tau}_{k,t}\|^2] $ in \eqref{eq_A2t_sum}. Then, \vspace*{-.7em}
\begin{align}
& A_{4,t} = \mae\left[\|\thetabf_t- \thetabf^{0}_{k,t}\|^2\right] + \sum_{\tau=1}^{J-1}\mae\left[\|\thetabf_t- \thetabf^{\tau}_{k,t}\|^2\right]  \nn\\
\!\!\!& \leq  2\sum_{\tau=0}^{J-1}\mae\left[\|\hat{\nbf}^{\text{dl}}_{k,t}\|^2\right]  + 2\sum_{\tau=1}^{J-1}\mae\bigg[\bigg\| \etat\!\sum_{\tau'=0}^{\tau-1}\!
\nabla F_{k}(\thetabf^{\tau'}_{k,t}; \Bc^{\tau'}_{k,t}) \bigg\|^2\bigg]   \nn\\
\!\!\!& \leq \frac{JD\sigma^2_{\text{d}}}
{|(\wbf^{\text{dl}}_t)^{\textsf{H}}\hbf_{k,t}|^2} + 2\eta_t^2\sum_{\tau=1}^{J-1}\!\tau\!\sum_{\tau'=0}^{\tau-1}\mae\left[\left\|\nabla F_{k}(\thetabf^{\tau'}_{k,t}; \Bc^{\tau'}_{k,t}) \right\|^2\right] \nn\\
\!\!\!& \stackrel{(a)}{=} \frac{JD\sigma^2_{\text{d}}}{|(\wbf^{\text{dl}}_t)^{\textsf{H}}\hbf_{k,t}|^2} + 2\eta_t^2\sum_{\tau=1}^{J-1}\!\tau\!\sum_{\tau'=0}^{\tau-1}\mae\left[\left\|\nabla F_{k}(\thetabf^{\tau'}_{k,t})\right\|^2\right]  \nn\\
\!\!\!&\quad + 2\eta_t^2\sum_{\tau=1}^{J-1}\!\tau\!\sum_{\tau'=0}^{\tau-1}\mae\left[\left\|\nabla F_{k}(\thetabf^{\tau'}_{k,t}; \Bc^{\tau'}_{k,t}) -  \nabla F_{k}(\thetabf^{\tau'}_{k,t})\right\|^2\right]  \nn\\
\!\!\!& \stackrel{(b)}{\leq}  \frac{JD\sigma^2_{\text{d}}}{|(\wbf^{\text{dl}}_t)^{\textsf{H}}\hbf_{k,t}|^2}\! + \! 2\eta_t^2\!\sum_{\tau=1}^{J-1}\!\tau\!\!\sum_{\tau'=0}^{\tau-1}\!\mae\!\left[\left\|\nabla F_{k}(\thetabf^{\tau'}_{k,t})\right\|^2\right]\!\! +\! 2\eta_t^2J^3\mu  \nn
\end{align}
where
$(a)$ and $(b)$ follow Assumption~\ref{assump_unbias}.
Let $A_{5,t}$ denote  $\sum_{\tau'=0}^{\tau-1}\mae[\|\nabla F_{k}(\thetabf^{\tau'}_{k,t})\|^2]$, which is bounded by\vspace*{-.7em}
\begin{align}
\!\!& A_{5,t}  \leq   2\!\sum_{\tau'=0}^{\tau-1}\!\mae\!\left[\left\| \nabla F_{k}(\thetabf^{\tau'}_{k,t}) -  \nabla F_{k}(\thetabf_t)\right\|^2\right]\!  \nn\\
\!\!\!\!\!& \qquad\;\; + 4J\mae\!\left[\| \nabla F(\thetabf_t)\|^2\right]
+ 4J\mae\left[\|   \nabla F_{k}(\thetabf_t) - \nabla F(\thetabf_t)\|^2\right]  \nn\\
\!\!& \leq \! 2L^2\!\!\sum_{\tau'=0}^{\tau-1}\!\mae\!\left[\| \thetabf^{\tau'}_{k,t}\! - \! \thetabf_t\|^2\right] \!+ \!4J\mae\left[\| \nabla F(\thetabf_t)\|^2\right]\! + \! 4J\delta. \label{eq_A5t}
\end{align}
Next, for $A_{3,t}$, based on Assumption~\ref{assump_bound_model},
we obtain
$A_{3,t} \leq  \sum_{k=1}^{K}\left|\Im\{\rho_{k,t}\}\right|\mae\left[\left\|\Delta\bar{\thetabf}_{k,t}\right\|
\left\|\nabla F(\thetabf_t)\right\|\right] \leq \nu\zeta\beta^{\text{im}}_{ t}$.
Combining the above bounds with \eqref{eq_A1},
we have \eqref{eq_A1_bound}. \endIEEEproof
\vspace*{-1em}
\section{Proof of Lemma~\ref{lemma2}}\label{appB}
\IEEEproof
For bounding $B_{1,t}$, we bound the first term in \eqref{eq_B1t_firststep},
denoted by $B_{2,t}$.
Specifically, we have\vspace*{-.5em}
\begin{align}
\!& B_{2,t} \leq   2\mae\bigg[\bigg\|\sum_{k=1}^{K}\rho_{k,t}\Delta\tilde{\thetabf}_{k,t}\bigg\|^2\bigg]+2\mae\bigg[\bigg\|\sum_{k=1}^{K}\rho_{k,t}\tilde{\nbf}^{\text{dl}}_{k,t}\bigg\|^2\bigg] \nn\\
\!& = 2\mae\bigg[\bigg\|\!\sum_{k=1}^{K}\!\Re\{\rho_{k,t}\}\Delta\tilde{\thetabf}_{k,t}\bigg\|^2\bigg]\!\! +\!  2\mae\bigg[\bigg\|\!\sum_{k=1}^{K}\!\Im\{\rho_{k,t}\}\Delta\tilde{\thetabf}_{k,t}\bigg\|^2\bigg] \nn\\
\!& \qquad + D\sigma^{2}_{\text{d}}\sum^{K}_{k=1}\frac{|\rho_{k,t}|^2}{|(\wbf^{\text{dl}}_t)^{\textsf{H}}
\hbf_{k,t}|^2}.
\label{eq_B2t_ini}
\end{align}
Let $B_{3,t}$ and $B_{4,t}$ respectively denote the first and
second terms in \eqref{eq_B2t_ini}.
Let $B_{3,t}$ and $B_{4,t}$ respectively denote the first and
second terms in \eqref{eq_B2t_ini}.
For $B_{3,t}$, we have\vspace*{-.7em}
\begin{align}
B_{3,t} & \leq 2(\beta^{\text{re}}_{ t})^2 \mae\bigg[\bigg\|\!\sum_{k=1}^{K}\! \frac{|\Re\{\rho_{k,t}\}|}{\beta^{\text{re}}_{ t}} \big\|\Delta\tilde{\thetabf}_{k,t}\big\|\bigg\|^2\bigg]  \nn\\
& \stackrel{(a)}{\leq} 2\beta^{\text{re}}_{ t}\sum_{k=1}^{K}\! |\Re\{\rho_{k,t}\}|\mae[\|\Delta\thetabf_{k,t}\|^2]
\label{eq_B3t}
\end{align}
where $(a)$ applies the Jensen's inequality.
We now bound the term $\mae\big[\big\|\Delta\thetabf_{k,t}\big\|^2\big]$ in \eqref{eq_B3t},
given by\vspace*{-.7em}
\begin{align}
&\mae\big[\big\|\Delta\thetabf_{k,t}\big\|^2\big] =   \eta_t^2 \mae\bigg[\bigg\|\sum_{\tau=0}^{J-1}\nabla F_{k}(\thetabf^{\tau}_{k,t}; \Bc^{\tau}_{k,t})\bigg\|^2\bigg]  \nn\\
& \stackrel{(a)}{=}  \eta_t^2 \sum_{\tau=0}^{J-1}\mae\bigg[\bigg\|\nabla F_{k}(\thetabf^{\tau}_{k,t}; \Bc^{\tau}_{k,t}) - \nabla F_{k}(\thetabf^{\tau}_{k,t})\bigg\|^2\bigg]  \nn\\
&\qquad + \eta_t^2 \mae\bigg[\bigg\|\sum_{\tau=0}^{J-1}\nabla F_{k}(\thetabf^{\tau}_{k,t})\bigg\|^2\bigg]   \nn\\
& \stackrel{(b)}{\leq} \eta_t^2J\mu + 2\eta_t^2\mae\bigg[\bigg\|\sum_{\tau=0}^{J-1}(\nabla F_{k}(\thetabf^{\tau}_{k,t}) - \nabla F_{k}(\thetabf_{t}))\bigg\|^2\bigg]  \nn\\
& \qquad + 2\eta_t^2 \mae\bigg[\bigg\|\sum_{\tau=0}^{J-1}\nabla F_{k}(\thetabf_{t})\bigg\|^2\bigg]
\label{eq_mae_delta}
\end{align}
where $(a)$ and $(b)$ are from Assumption~\ref{assump_unbias}.
Let $B_{5,t}$ and $B_{6,t}$ respectively denote the second and third
terms in \eqref{eq_mae_delta}.
For $B_{5,t}$, we have
\begin{align}
B_{5,t} & \leq 2\eta_t^2J\sum_{\tau=0}^{J-1} \mae\left[\|\nabla F_{k}(\thetabf^{\tau}_{k,t}) - \nabla F_{k}(\thetabf_{t})\|^2\right]  \nn\\
& \stackrel{(a)}{\leq} 2\eta_t^2JL^2\sum_{\tau=0}^{J-1} \mae\left[\|\thetabf^{\tau}_{k,t} - \thetabf_{t}\|^2\right] \nn\\
& \stackrel{(b)}{\leq}\frac{(1-Q_t)^2}{L^2Q_t}\mae[\|\nabla F(\thetabf_t)\|^2]+ \frac{(1-Q_t)D\sigma^2_{\text{d}}}{2Q_t|
(\wbf^{\text{dl}}_t)^{\textsf{H}}\hbf_{k,t}|^2}   \nn\\
&\qquad + \frac{(1-Q_t)^2}{2L^2Q_t}(4\delta+\mu)
\label{eq_B5t}
\end{align}
where $(a)$ follows the $L$-smoothness of $F_k(\cdot)$ in Assumption~\ref{assump_smooth},
and $(b)$ applies the bounds on $A_{4,t}$.
For $B_{6,t}$, we have
\begin{align}
& B_{6,t}  = 2\eta_t^2J^2 \mae\left[\|\nabla F_k(\thetabf_t) \|^2\right]\nn\\
& \leq   4\eta_t^2J^2  \mae\left[\|\nabla F_k(\thetabf_t)\! -\! \nabla F(\thetabf_t)\|^2\right]\! +\! 4\eta_t^2J^2 \mae\left[\|\nabla F(\thetabf_t)\|^2\right]\nn\\
&  \stackrel{(a)}{\leq}  4\eta_t^2J^2\delta + 4\eta_t^2J^2 \mae\left[\|\nabla F(\thetabf(t))\|^2\right] \label{eq_B6t}
\end{align}
where $(a)$ follows Assumption~\ref{assump_bound_diverg}.
Next, for  $B_{4,t}$, based on Assumption~\ref{assump_bound_model},
 we obtain
 $B_{4,t} \leq 2\mae[\|\!\sum_{k=1}^{K}\!|\Im\{\rho_{k,t}\}|\|\Delta\tilde{\thetabf}_{k,t}\|\|^2] \leq 2\nu^2(\beta^{\text{im}}_{ t})^{2}$.
Combining the above bounds with \eqref{eq_B2t_ini}\eqref{eq_B1t_firststep},
we have \eqref{eq_B1t_lemma2}.
\endIEEEproof

\section{Proof of Proposition~\ref{thm:convergence}}\label{appC}
\IEEEproof
We apply Lemmas~\ref{lemma1} and \ref{lemma2} to \eqref{theta_diff}.
Let $Y_t \triangleq \frac{\eta_tJ}{2}\big(4V_t(\beta^{\text{re}}_{ t})^2 - 1\big)$. Using the definition of $V_t$, we have
$Y_t = \frac{\eta_tJ}{2Q_t}\big(4(1-Q_t)(\beta^{\text{re}}_{ t})^2 + 4\sqrt{1-Q_t}(\beta^{\text{re}}_{ t})^2  -  Q_t \big)$.
Letting $x\triangleq\sqrt{1-Q_t}$,
we have
\begin{align}
Y_t = \frac{\eta_tJ}{2Q_t}\left(\left(4(\beta^{\text{re}}_{ t})^2 + 1\right)x^2 + 4(\beta^{\text{re}}_{ t})^2x - 1\right). \nn
\end{align}
We first obtain $Q_t>0$ for $\eta_tJ<\frac{1}{2L}$.
Next, the positive root of equation $\big(4(\beta^{\text{re}}_{ t})^2 + 1\big)x^2 + 4(\beta^{\text{re}}_{ t})^2x - 1 = 0$ is given by
\begin{align}
x^o & = \frac{-4(\beta^{\text{re}}_{ t})^2 + \sqrt{16(\beta^{\text{re}}_{ t})^4 + 16(\beta^{\text{re}}_{ t})^2 + 4}}{8(\beta^{\text{re}}_{ t})^2 + 2}  \nn\\
& = \frac{1}{4(\beta^{\text{re}}_{ t})^2 + 1}.  \nn
\end{align}
Thus, for $x\geq x^o$, we have $\big(4(\beta^{\text{re}}_{ t})^2 + 1\big)x^2 + 4(\beta^{\text{re}}_{ t})^2x - 1 \geq 0$.
Re-express $x\geq x^o$ as $\eta_tJ\geq\frac{1}{2L(4(\beta^{\text{re}}_{ t})^{2}+ 1)}$.
Note that $\beta^{\text{re}}_{ t} = \sum_{k=1}^{K}\!|\Re\{\rho_{k,t}\}| \geq |\Re\{\sum_{k=1}^{K}\rho_{k,t}\}|=1$,
leading to $\frac{1}{2L(4(\beta^{\text{re}}_{ t})^{2}+ 1)} < \frac{1}{2L}$.
Based on these, we have $Y_t\geq 0$ for $\frac{1}{2L(4(\beta^{\text{re}}_{ t})^{2}+ 1)}\le \eta_tJ<\frac{1}{2L}$.

Let $Z_t \triangleq H(\wbf^{\text{dl}}_t, \rhobf_t)
+ C_{t}$.
We have $Z_t>0$ due to $H(\wbf^{\text{dl}}_t, \rhobf_t) > 0$ and
$C_{t} > 0$.
Then, after combining
\eqref{eq_A1_bound}, \eqref{eq_B1t_lemma2}, and \eqref{theta_diff},
for $\frac{1}{2L(4(\beta^{\text{re}}_{ t})^{2}+ 1)}\le \eta_tJ<\frac{1}{2L}$,
we have
\begin{align}
&\mae[F(\thetabf_{t+1})]\!-\!F^{\star} \le\mae[F(\thetabf_t)]\!-\!F^{\star}\!+\!Y_t\mae[\|\nabla F(\thetabf_t)\|^2]\!+\!Z_t  \nn\\
& \stackrel{(a)}{\leq} \mae[F(\thetabf_t)]\!-\!F^{\star}\!+\!L^2Y_t\mae[\left\|\thetabf_t - \thetabf^{\star}\right\|^2]\!+\!Z_t \nn\\
& \stackrel{(b)}{\leq} \mae[F(\thetabf_t)]\!-\!F^{\star}\!+\!\frac{2L^2Y_t}{\lambda}(\mathbb{E}[F(\thetabf_t)] - F^{\star})\!+\!Z_t \label{eq_diff_upper_boundHC}
\end{align}
where $(a)$ and $(b)$ follow from \eqref{eq_global_local_equation} and  the $L$-smoothness and  $\lambda$-strong-convexity of $F_k(\cdot)$ in   Assumption~\ref{assump_smooth}, respectively.
Summing up both sides of \eqref{eq_diff_upper_boundHC} over $t\in\Tc$ and  rearranging the terms,
we have \eqref{eq_thm1}.
\endIEEEproof
\end{appendices}

\bibliographystyle{IEEEtran}
\bibliography{Refs}

\end{document}